\documentclass[acmlarge,screen]{acmart}

\AtBeginDocument{%
  }

\setcopyright{acmcopyright}
\copyrightyear{2018}
\acmDOI{XXXXXXX.XXXXXXX}

\acmJournal{TOSEM}
\usepackage{booktabs} 

\usepackage{graphicx}
\usepackage{subfiles}
\usepackage{array}
\usepackage{amsmath}
\usepackage[ruled,linesnumbered]{algorithm2e}
\usepackage{multirow}
\usepackage{subfigure}
\usepackage{caption}
\usepackage{flushend}
\usepackage{balance}
\usepackage{diagbox}
\usepackage{colortbl}
 \usepackage{enumitem}
 \usepackage{balance}
 \usepackage{makecell}
 \usepackage{float}
\usepackage{fancyhdr}
\usepackage{threeparttable}

\usepackage{url}
\usepackage{bm}
\usepackage{comment}
\usepackage{graphicx}
\usepackage{tcolorbox}
\usepackage{subfigure}

\usepackage{listings}

\newcommand{\tool}{\textsc{SynAT}}

\newcommand{\andrelation}{\textsc{And}}
\newcommand{\orrelation}{\textsc{Or}}
\newcommand{\parentrelation}{\textsc{AchievedBy}}

\begin{document}




\title{{\tool}: Enhancing Security Knowledge Bases via Automatic Synthesizing Attack Tree from Crowd Discussions}

\author{Ziyou Jiang}
\orcid{0000-0003-1182-143X}
\affiliation{%
    \institution{State Key Laboratory of Intelligent Game}
  \state{Beijing}
  \country{China}
}
\affiliation{%
    \institution{Science and Technology on Integrated Information System Laboratory, Institute of Software Chinese Academy of Sciences}
  \state{Beijing}
  \country{China}
}
\affiliation{%
  \institution{University of Chinese Academy of Sciences}
  \state{Beijing}
  \country{China}
}
\email{ziyou2019@iscas.ac.cn}
\authornote{Both authors contribute equally.}

\author{Lin Shi}
\orcid{0000-0003-1476-7213}
\affiliation{%
  \institution{School of Software, Beihang University}
  \state{Beijing}
  \country{China}
}
\email{shilin@buaa.edu.cn}
\authornotemark[1]

\author{Guowei Yang}
\orcid{0000-0002-1404-4560}
\affiliation{%
  \institution{School of Information Technology and Electrical Engineering, The University of Queensland}
  \city{Brisbane}
  \state{Queensland}
  \country{Australia}
}
\email{guowei.yang@uq.edu.au}

\author{Xuyan Ma}
\orcid{0000-0002-8514-2336}
\affiliation{%
    \institution{State Key Laboratory of Intelligent Game}
  \state{Beijing}
  \country{China}
}
\affiliation{%
    \institution{Science and Technology on Integrated Information System Laboratory, Institute of Software Chinese Academy of Sciences}
  \state{Beijing}
  \country{China}
}
\affiliation{%
  \institution{University of Chinese Academy of Sciences}
  \state{Beijing}
  \country{China}
}
\email{maxuyan2021@iscas.ac.cn}

\author{Fenglong Li}
\orcid{0000-0002-0030-8166}
\affiliation{%
  \institution{Huawei Cloud Computing Technologies CO., LTD.}
  \state{Beijing}
  \country{China}
}
\email{lifenglong2@huawei.com}

\author{Qing Wang}
\orcid{0000-0002-2618-5694}
\affiliation{%
    \institution{State Key Laboratory of Intelligent Game}
  \state{Beijing}
  \country{China}
}
\affiliation{%
    \institution{Science and Technology on Integrated Information System Laboratory, Institute of Software Chinese Academy of Sciences}
  \state{Beijing}
  \country{China}
}
\affiliation{%
  \institution{University of Chinese Academy of Sciences}
  \state{Beijing}
  \country{China}
}
\email{wq@iscas.ac.cn}
\authornote{Corresponding author.}


\begin{abstract}
{Cyber attacks have become a serious threat to the security of software systems. Many organizations have built their security knowledge bases 
to safeguard against attacks and vulnerabilities.
However, due to the time lag in the official release of security information, these security knowledge bases may not be well maintained, and using them to protect software systems against emergent security risks can be challenging.
On the other hand, the security posts on online knowledge-sharing platforms contain many crowd security discussions
and the knowledge in those posts can be used to enhance the security knowledge bases.
This paper proposes {\tool}, an automatic approach to synthesize attack trees from crowd security posts. Given a security post, {\tool} first utilize the Large Language Model (LLM) and prompt learning to restrict the scope of sentences that may contain attack information; then it utilizes a transition-based event and relation extraction model to extract the events and relations simultaneously from the scope; finally, it applies heuristic rules to synthesize the attack trees with the extracted events and relations. An experimental evaluation is conducted on 5,070 Stack Overflow security posts, and the results show that {\tool} outperforms all baselines in both event and relation extraction, and achieves the highest tree similarity in attack tree synthesis. 
Furthermore, {\tool} has been applied to enhance HUAWEI's security knowledge base as well as public security knowledge bases CVE and CAPEC, which demonstrates {\tool}'s {practicality}.
}
\end{abstract}



\maketitle
\section{Introduction}


{Cyber attacks have become a serious threat to the security of software systems~\cite{DBLP:journals/compsec/XiongL19}, and some attacks from malicious users may result in millions of dollars of losses {in today’s businesses}~\cite{article}.
Many research institutions and organizations have built
their security knowledge base to help safeguard against attacks and vulnerabilities.
For example,
\textbf{Common Vulnerabilities and Exposure (CVE)}~\cite{CVE} is one of the widely-used public security knowledge databases that identifies each vulnerability with a unique name and standard description.
Security practitioners can retrieve the relevant vulnerability from this database to patch the vulnerable projects and enhance their security.
Another example is \textbf{Huawei Cloud Computing Technologies}, which is a Fortune Global 500 company in 2023, has set up a security knowledge base that consists of 241 attack trees and carries important information such as attack description, attack case, and possible mitigation in terms of STRIDE \cite{STRIDE} threat model. 
The security knowledge bases provide developers and maintainers with code design specifications and threat modeling analysis of {their products}.

However, due to the time lag in the official release of security information, the knowledge base has not been maintained regularly, and the time lag may be utilized by attackers to deploy exploits, such as zero-day attacks. 
For CVE, Table \ref{tab:CVE_disclosure} shows the disclosure time of the vulnerabilities and the created time of posts. We can see that some of the posts were created over 400 days earlier than the disclosure, and the time lags may be easily utilized by attackers.
For HUAWEI's database, 77$\%$ of the attack trees have not been updated within the last four years. 
Some latest well-known attacks, such as \emph{format-string vulnerability} have not been included in the knowledge base.  
Therefore, it is challenging for the company to protect its software systems against emergent security risks.}

\begin{table}[htbp]
  \centering
  \tiny
    \vspace{-0.3cm}
  \caption{{The vulnerabilities disclosed in CVE that are first proposed in Stack Overflow's security posts.}}
  \vspace{-0.3cm}
     \resizebox{\columnwidth}{!}{\begin{tabular}{|c|l|l|l|}
     \hline
      \textbf{CVE-ID} & \multicolumn{1}{c|}{\textbf{Time of Disclosure}} & \multicolumn{1}{c|}{\textbf{Corresponding Security Post}} & \multicolumn{1}{c|}{\textbf{Time of Post Created}} \\
      \hline
    CVE-2011-1271 & 05-10-2011 & \#2135509/bug-only-occurring-when-compile-optimization-enabled & 01-25-2010 \textbf{(-470 days)}\\
    CVE-2012-5633 & 03-12-2013 & \#7933293/why-does-apache-cxf-ws-security-implementation-ignore-get-requests & 10-28-2011 \textbf{(-501 days)}\\
    CVE-2013-3350 & 07-10-2013 & \#17351214/cf10-websocket-p2p-can-invoke-any-public-functions-in-any-cfc & 06-27-2013 \textbf{(-13 days)}\\
    CVE-2015-3833 & 10-01-2015 & \#24625936/getrunningtasks-doesnt-work-in-android & 07-08-2014 \textbf{(-450 days)}\\
    CVE-2015-3198 & 07-21-2017 & \#30028346/with-trailing-slash-in-url-jsp-show-source-code & 05-04-2015 \textbf{(-809 days)}\\
    \hline
    \end{tabular}%
    }
  \label{tab:CVE_disclosure}%
\end{table}%

\begin{figure}[htbp]
\centering
\subfigure[Security post \#36817325.]{\includegraphics[width=0.425\columnwidth]{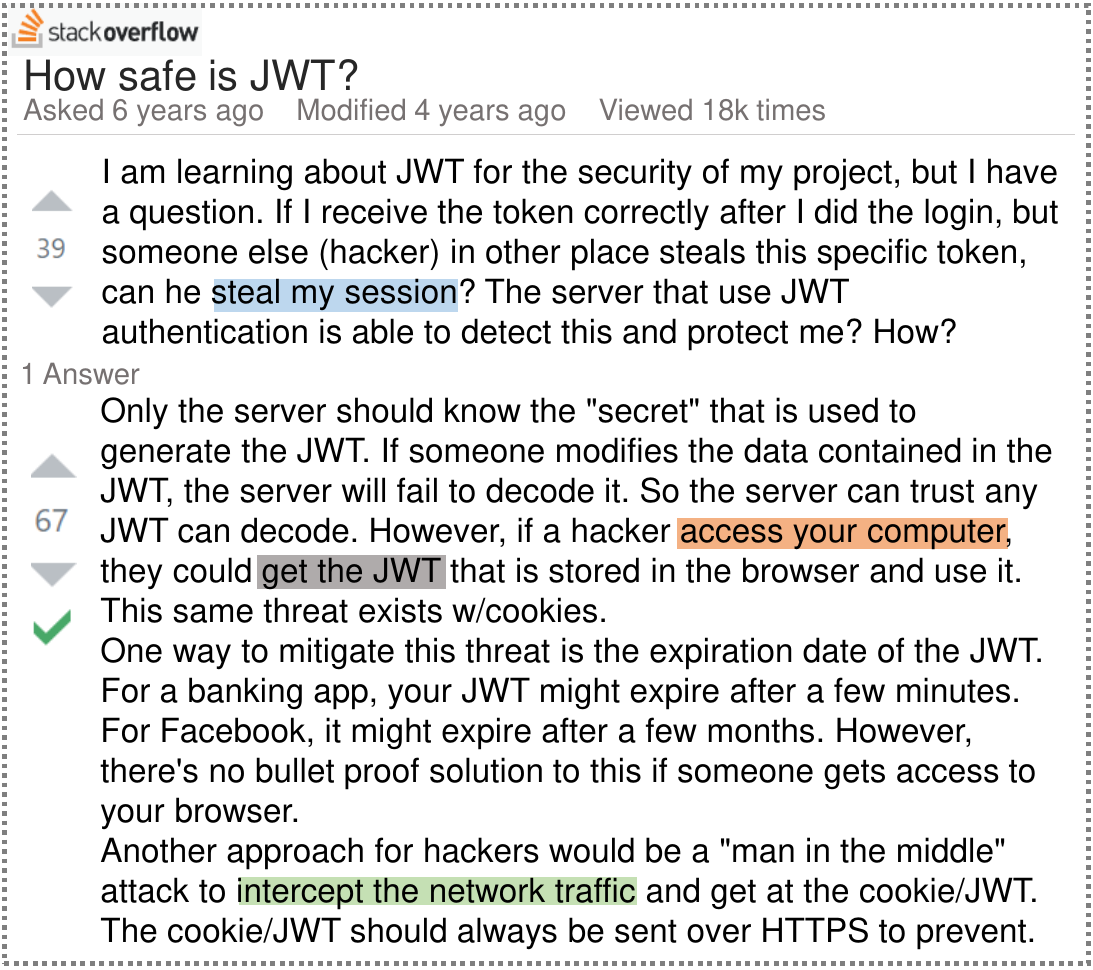}}
\hspace{-0.1cm}
\subfigure[Synthesized attack tree.]{\includegraphics[width=0.3\columnwidth]{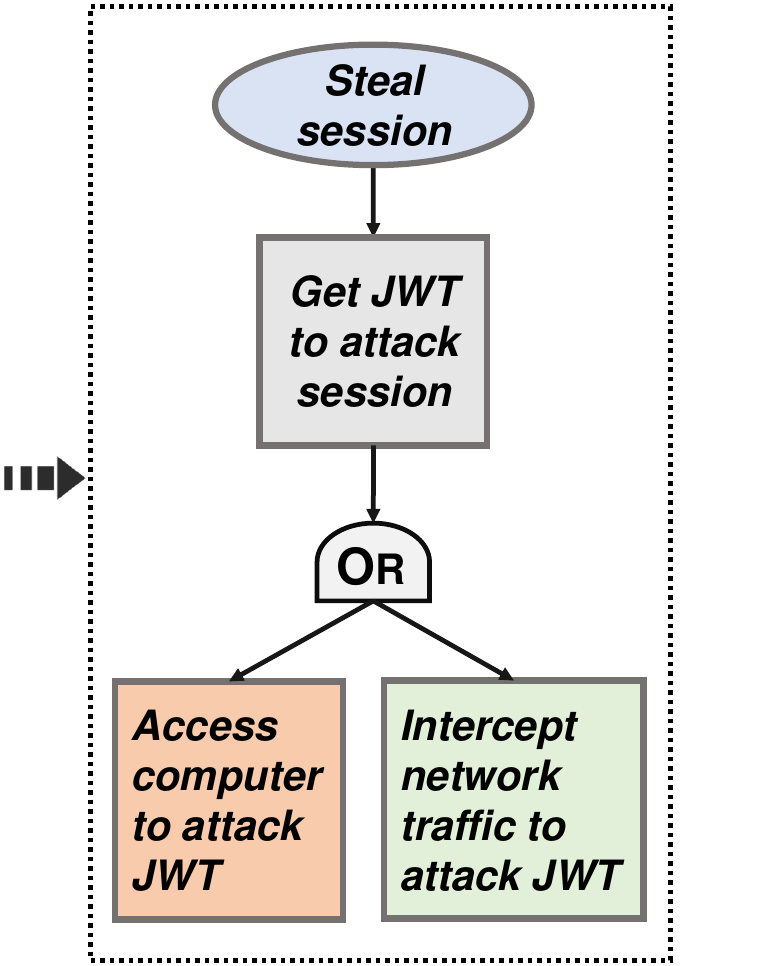}}
\vspace{-0.2cm}
\caption{The motivation example of attack tree synthesizing in security post \#35817325. 
}
\vspace{-0.4cm}
\label{fig:motivation}
\end{figure}

Nowadays, developers tend to leverage online knowledge-sharing platforms, such as Stack Overflow, GitHub, etc., to discuss their security concerns with other developers about possible attacks, and ask for suggestions to develop secure software. 
{These security discussions can be obtained from the security posts in Stack Overflow, and the GitHub Issues for different projects. 
{Pan et al.~\cite{DBLP:conf/sigsoft/PanZC0BHLH22} indicate that developers conduct the security discussions and they may pertain to the attacks that {have not ever been officially reported}}. {Identifying such emerging attacks on time will help prevent losses}.
In addition, due to the diversity and practical nature of Stack Overflow, the extracted information 
will enhance security knowledge bases for various IT organizations.
}
Fig. \ref{fig:motivation} (a) shows an example, where developers discuss attacking and securing JWT in Stack Overflow. 
In particular, the questioner is concerned that hackers may steal the tokens and access their sessions. In the accepted answer, two possible attack methods are enumerated: access the computer or intercept the network traffic. If any of these attack methods succeeded, hackers could get the JWT and finally access the session data.
{According to this discussion, we can {potentially} synthesize an attack tree, where the attack goal ``\textit{\textbf{steal session}}'' could be achieved via two attack methods, as shown in Fig. \ref{fig:motivation} (b).} {The synthesized attack trees are invaluable for the {software security community}. 
}

However, directly synthesizing these attack trees that are buried in the textual security post is non-trivial, since the complex attack trees are highly abstract and reasonably inferred from multiple attack events that are described in the textual security discussions. 
To bridge the gap, we intuitively accommodate the general event and relation extraction to the security posts, then match the extracted elements to attack trees. 
In this paper, we propose {\tool}, an automatic approach to {synthesize} attack trees from security posts of Stack Overflow with transition-based event and relation extraction. 
Specifically, 
{{\tool} first restricts the scope of sentences that may contain the attack information with Large Language Model (LLM) and with the auto-generated prompt, which has the State-of-the-Art (SOTA) performances on multiple natural language process tasks~\cite{petroni-etal-2019-language}.}
Then, it utilizes a transition-based joint event and relation extraction model to extract the attack events and their relations simultaneously from the scoped sentences.
Finally, it applies a set of heuristic rules to synthesize the attack trees with the extracted events and relations.

To evaluate the performance of {\tool} in attack tree synthesizing, 
{we conduct experiments on both Stack Overflow and GitHub, with 5,070 security posts, and 2,350 GitHub issue reports (IRs) labeled from the GHArchive~\cite{GHArchive}, which is a large-scale IR-based security dataset collected from 2015. 
Then, we compare {\tool} with multiple representative baselines on attack tree synthesizing, event extraction, and relation extraction.}
The results show that, 
{{\tool} achieves the highest tree similarity in attack tree synthesizing, with 10.24\% Average Hamming Distance (AHD) and 7.93\% Tree-edit Distance Similarity (TEDS). {\tool} also outperforms all baselines in both event and relation extraction, with 80.93\% and 87.81\% F1 scores, respectively.} 
Furthermore, {\tool} has been practically applied to enhance the public security knowledge bases, i.e., CVE~\cite{CVE} and CAPEC~\cite{CAPEC1}, as well as HUAWEI's security knowledge base.
The major contributions of this paper are summarized as follows:

\begin{itemize}[leftmargin=*]
    \item
    \textbf{Technique}: {\tool}, {an automated approach to synthesize attack trees based on LLM and joint event and relation extraction.} 
    To the best of our knowledge, this is the first work on automatically synthesizing attack trees from the crowd security posts of a knowledge-sharing platform.
    \item 
    \textbf{Evaluation}: An experimental evaluation of the performance of {\tool} against state-of-the-art baselines, which shows that {\tool} outperforms all baselines. 
    \item \textbf{Application}: 
    {Practical applications in enhancing HUAWEI's security knowledge base and public security knowledge base, which demonstrate {\tool}'s {practicality}.}
    \item \textbf{Data}:
    {A public release of the dataset with 1,354 attack trees, and source code~\cite{model_data} to facilitate the replication of our study and its application in extensive contexts.}
\end{itemize}

{In the rest of the paper, Section 2 illustrates the background and motivation. Section 3 presents the details of our approach. Section 4 sets up the experiments. Section 5 describes the experimental results and analysis. Section 6 describes the application study. Section 7 presents the discussion and threats to validity. Section 8 discusses the related work, and Section 9 concludes this paper.}

\section{Preliminaries}

\subsection{Attack Tree}
\label{background:tree}

\begin{figure}[htbp]
\centering
\includegraphics[width=0.6\columnwidth]{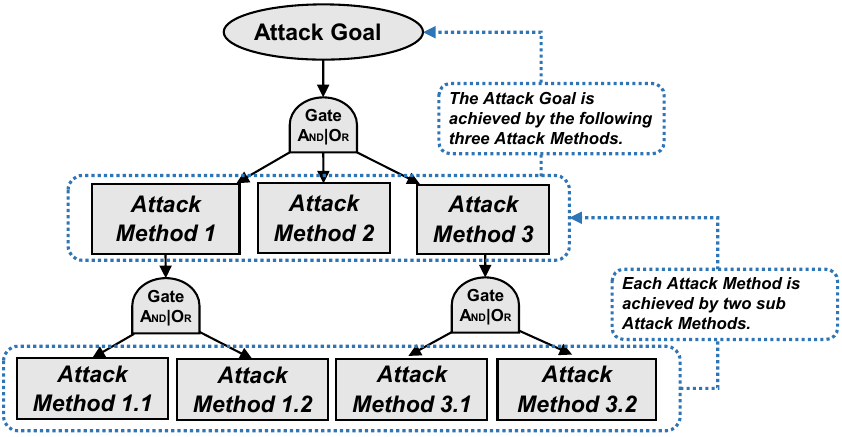}
\caption{The meta-framework of attack trees.}
\label{fig:attacktree}
\end{figure}

{Our study aims to synthesize attack trees from security posts. 
An attack tree is a tree-structured conceptual diagram, which demonstrates
attack behaviors of invading a system in a formal way~\cite{7943455}. Since attack trees can enumerate the possible attacks to exploit a system~\cite{schneier1999attack}, they {have been} widely used in security research and practice, such as threat modeling \cite{DBLP:journals/compsec/XiongL19}, to boost the target systems' security~\cite{jiang2014energy,DBLP:conf/sec/JhawarKMRT15,10.1093/logcom/exs029,DBLP:conf/smartcloud/Straub20}.
Typically, an attack tree is composed of one \textbf{{{Attack Goal}}}, the \textbf{{Attack Methods}} that can achieve the attack goal, and the \textbf{{Relations}} between attack methods (i.e. \textbf{{\andrelation}} and \textbf{{\orrelation}}), as shown in Fig. \ref{fig:attacktree}.
}

\subsection{Motivation}\label{sec:motivation}

Directly extracting attack trees from security posts (Security Post $\rightarrow$ Attack Tree) is extremely challenging since the complex attack trees are highly abstract and reasonably inferred from multiple attack events that are described in the textual security posts.
Therefore, we reduce the abstract level by reformulating the attack tree extraction task into a two-step extraction, i.e., we first extract attack events and their relations from security posts and then synthesize the attack trees from these attack events (Security Post $\rightarrow$ Attack Events $\rightarrow$ Attack Tree).

\subsubsection{Definition of Attack Event.}
Event Extraction is an important task in Natural Language Process~\cite{DBLP:conf/acl/LiaoG10}. An event refers to a specific occurrence of facts that happen in a certain time and a certain place, which can usually be described as a change of state~\cite{DBLP:conf/lrec/DoddingtonMPRSW04}. 
Extracted Events contain one or more \textbf{triggers} (i.e., the main words or phrase that most clearly expresses an event occurrence) and \textbf{arguments} (i.e., roles involved in an event, describing the main components of the event). For example, the triggers for banking events could be the phrase `Transfer Money', and the types of argument could be `Beneficiary' and `Recipient'.
Event extraction requires identifying the event, classifying triggers, identifying arguments, and judging the argument type. In this paper, we define the attack event inspired by the previous work~\cite{7943455}. The \textbf{{Attack Event}} is composed of one \textbf{{Trigger}}, and two types of arguments (i.e., \textbf{Target} and \textbf{Instrument}).

\begin{itemize}[leftmargin=*]
    \item \textbf{Trigger}: Main verbs indicating security attacks.
    \item \textbf{{Target}}: Objects aimed by attackers to achieve attacks.
    \item \textbf{{Instrument}}: Tools utilized by attackers to achieve attacks.
\end{itemize}




\subsubsection{Definition of Relation.}
The relation extraction is a task to predict the relation between a pair of events. Ning et al. \cite{ning-etal-2018-improving} extract the temporal relations between events, such as \textsc{Before} and \textsc{After}. Glavas et al. \cite{DBLP:conf/lrec/GlavasSMK14} analyze the {structural} relations between events, which are mainly classified as \textsc{Parent-Child} and \textsc{Child-Parent}.
In this study, we adapt the identification of attack tree edges to the relation extraction task. We define three types of relations that are useful for constructing attack trees as follows:

\begin{itemize}[leftmargin=*]
    \item \textbf{\andrelation}: Two events have an {\andrelation} relation if they both have to be satisfied to achieve another event, corresponding to the {\andrelation} relation in the attack tree.
    \item \textbf{\orrelation}: Two events have an {\orrelation} relation if each of them can achieve an event independently, corresponding to the {\orrelation} relation in the attack tree.
    \item \textbf{\parentrelation}: {An event is {\parentrelation} of another event, if the former event can be achieved by satisfying the latter event itself or together with other events.} This relation corresponds to the parent-child edge in the attack tree, which means the attack event can be the parent node of other events if it is achieved by these events.
\end{itemize}

\begin{figure}[htbp]
\centering
\includegraphics[width=0.6\columnwidth]{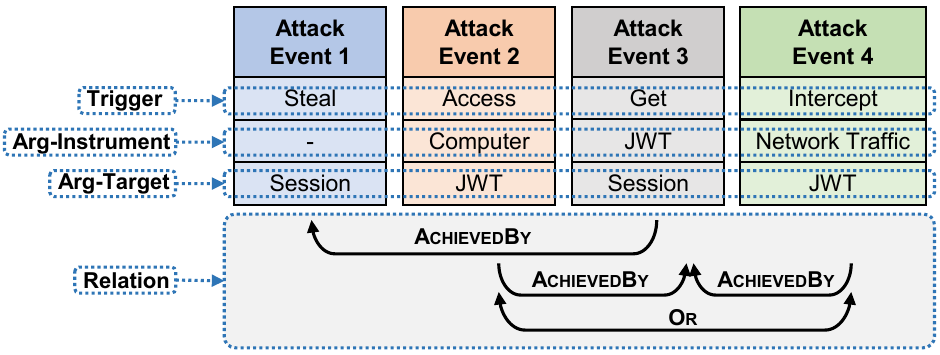}
\caption{The extracted attack events and relations of motivation example in Fig. \ref{fig:motivation}.
}
\label{fig:eventrelation}
\end{figure}




Fig. \ref{fig:eventrelation} shows an example of attack events and relations from the security post illustrated in Fig. \ref{fig:motivation}(a). Four attack events and their relations can be identified from the post. 
Based on the events and relations, we can synthesize the attack tree in Fig. \ref{fig:motivation}(b).

\subsection{Transition-based Extraction {Framework}}\label{sec:transtion_motivation}

Recent research evidenced that the transition-based information extraction framework is efficient for extracting information that can be described as a change of states. For example, the transition-based extraction model has achieved state-of-the-art performances on the entity and relation extraction~\cite{DBLP:conf/ijcai/WangZC018}, dependency parsing~\cite{DBLP:conf/aaai/WangCGL18}, and semantic role labeling~\cite{DBLP:conf/acl/ChoiP11a}.
{The basic idea of the transition-based framework is transforming the task of predicting a graph from a textual document, into predicting an action sequence of a state machine that produces the graph}~\cite{DBLP:journals/coling/ZhangC11,DBLP:conf/emnlp/AstudilloBNBF20}.
In particular, a transition-based framework has two key components: (1) transition \textbf{states} and (2) a set of transition \textbf{actions},  
where the state is used to record incomplete prediction results, and the action is used to control the transition between states.

{The transition-based extraction framework has two major advantages compared with the other event and relation extraction models, as well as the LLMs that have been proposed recently:
(1) Since the transition-based event and relation extraction only requires traversing the document once to obtain all the events and relations, it has a time complexity of $\mathcal{O}(n)$, which is faster than all the other models. 
(2) The transition-based model has high accuracy when specifying the types of events and relations, and more details are shown in Section \ref{sec:results}.}

In this study, since the attack events can also be described as a change of states, we employ the transition-based extraction framework to facilitate the attack event and relation extraction by newly designing a transition system, and jointly performing event extraction and relation extraction as a single task.

\section{Approach}

\begin{figure*}[t]
\centering
\includegraphics[width=\textwidth]{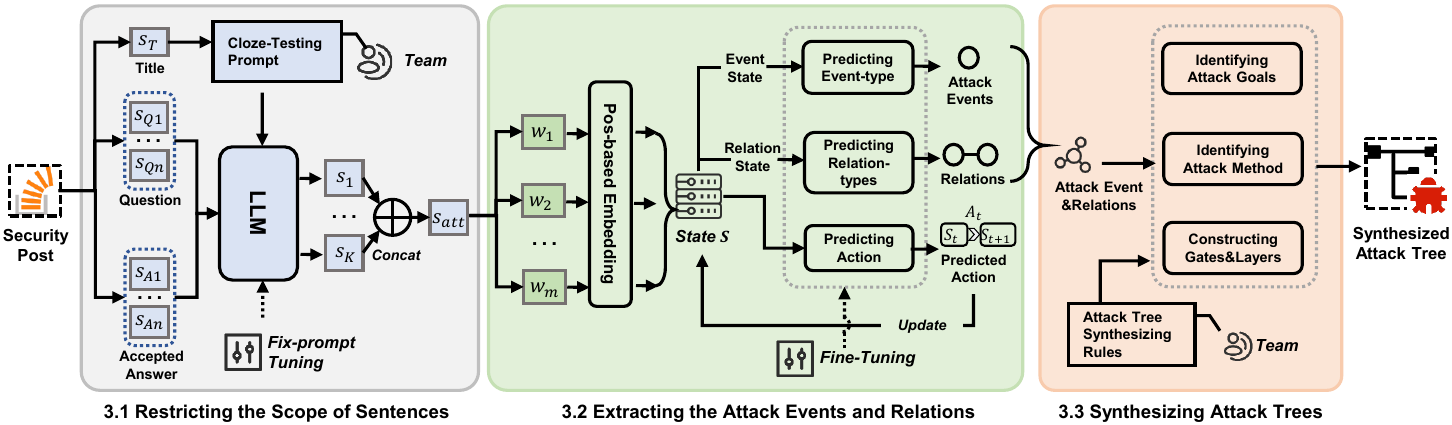}
\caption{The architecture of {\tool}.}
\label{fig:model}
\end{figure*}

The overview of {\tool} is illustrated in Fig. \ref{fig:model}, which consists of four major steps. 
{First, we restrict the scope of sentences from the security discussions with LLM, which may contain the attack events and relations.
Second, we introduce a transition-based event and relation extraction model to extract attack events and relations simultaneously from the restricted scope.
Third, we synthesize the attack tree with manually designed rules based on the extracted events and relations.}

\subsection{Restricting the Scope of Sentences} \label{sec:limiter}

A security post $\mathcal{P}$
consists of a \textit{Title} $S_{T}$, \textit{Question} $[S_{Q1},S_{Q2},...,S_{Qn}]$ and \textit{Accepted Answer} $[S_{A1},S_{A2},...,S_{An}]$, where the \textit{Question} and the \textit{Accepted Answer} are all composed of a sequence of sentences.
Among these sentences in \textit{Question} and \textit{Answer}, only a relatively small number of sentences are relevant to the attack events and their relations. 
To reduce the impact of non-relevant sentences, we restrict the sentences that are more likely to contain the attack events and their relations.

To achieve that, we utilize the LLM to restrict the scope of sentences and use Prompt Learning for tuning the LLM~\cite{schick2021few}. The paradigm LLM+Prompt has achieved state-of-the-art performances on various natural language process (NLP) tasks~\cite{liu2023pre}. Compared with the previous Pre-train+Fine-tuning paradigm, prompt learning can fully utilizes the resources of models with its Prompt Template~\cite{houlsby2019parameter}, and bridge gaps between inputs and LLMs.


\subsubsection{LLM-based Scope Restriction.}\label{sec:title-embedding}

To restrict the scope of sentences with the prompt learning, we design a template based on the \textbf{Cloze-testing}, which is a widely-used prompt template in text extraction task~\cite{petroni-etal-2019-language}. 
We compare three widely-used templates, i.e., \textbf{Cloze Template}, \textbf{Prefix Template}, and \textbf{Template w/o Title}, and choose the best-performed one. More details will be introduced in Section \ref{sec:limiter}).
Given the security post $\mathcal{P}$, the template for restricting the scope is shown as follows:
\begin{center}
\vspace{-0.2cm}
\small
\begin{tcolorbox}[colback=white,
                  colframe=black,
                  width=\columnwidth,
                  arc=1mm, auto outer arc,
                  boxrule=0.4pt,
                  left=0.1pt,
                  right=0.1pt,
                  top=0.1pt,
                  bottom=0.1pt,
                colbacktitle=white!80!gray, coltitle=black, 
                title={\textbf{Cloze-Template Prompt for LLM-based Scope Restriction}}
                 ]
\textit{\textbf{Cloze-Testing:} From the following security post \{Question, Accepted\_Answer\} (Title is \{Title\} to summarize the main topic of security post), \textbf{[Y]} are restricted sentences that may contain the attack trees.} 
\textit{(Omit the definition of attack trees.)}

\textit{\textbf{Task:} Please predict the token \textbf{[Y]} with $K$ restricted sentences from the security post's question and answers to make the cloze-testing complete. The format of the output is the bullet list with \textbf{Sentence 1} to \textbf{Sentence $K$}.}
\end{tcolorbox}
\end{center}
where the \{Question, Accepted\_Answer\} indicates the inputted sentences in \textit{Question and Accepted Answer}. 
Since the \textit{\{Title\}} of the security post summarizes the main topic of the security post, we incorporate it in the template to restrict the scope more accurately.
The token \textbf{\textit{[Y]}} is the output of restricted sentences that may contain the attack trees, and we ask the LLM to predict the sentences in this token to make the close testing complete.
We choose the generative GPT-3~\cite{DBLP:conf/nips/BrownMRSKDNSSAA20} as the LLM, which outperforms other LLMs on the text generation tasks. We utilize the GPT-3 to output the restricted $K$ sentences with the \textit{[Y]} token as $[S_1,...S_K]$, then concatenate them into one single sentence as the sentence-scope $S_{att}$.

\subsubsection{Fix-prompt Tuning}
To train the LLM, we apply the Fix-prompt Tuning~\cite{DBLP:journals/corr/abs-2103-10385}, which is a typical training method for the manually designed prompt templates:
\begin{equation}
    \mathcal{L}_{sco}=Cr(y_{sco},S_{att})
\end{equation}
where function $Cr$ indicates the CRINGE loss~\cite{DBLP:journals/corr/abs-2211-05826}, which is specifically used for training generative LLMs; $y_{sco}$ is the ground-truth label for the sentences that contain the attack tree.
We train the LLM with the loss $\mathcal{L}_{sco}$ until it achieves the training convergence.

\subsubsection{Comparison of LLMs}

To select the best LLM for restricting the scopes of sentences, we compare the performances with four representative LLMs on restricting the sentence scopes, i.e., \textbf{Albert}~\cite{DBLP:conf/iclr/LanCGGSS20}, \textbf{T5}~\cite{DBLP:journals/jmlr/RaffelSRLNMZLL20}, and \textbf{GPT-3}~\cite{DBLP:conf/nips/BrownMRSKDNSSAA20}, and \textbf{ChatGPT}~\cite{LLMBackground}.
Compared with other LLMs, these models achieve the highest performances on different NLP tasks.
We utilize the same fix-prompt tuning method to train the Albert, T5, and GPT-3. Since ChatGPT has not opened the fine-tuning interface, we choose in-context learning (ICL) to optimize it. According to the previous works~\cite{DBLP:conf/emnlp/MinLHALHZ22,DBLP:conf/emnlp/SinghCG0NV23}, the ICL is the effective few-shot strategy that can enable the ChatGPT on various NLP tasks, and we find that 5-shot ICL can achieve the highest performance on sentence restricting.  
To measure the performances, we utilize three metrics to analyze the performances of restricting the sentences, i.e., \textbf{Precision} (\textbf{Pre.}), \textbf{Recall} (\textbf{Rec.}), and \textbf{F1-measure} (\textbf{F1}):
\begin{equation}
    Pre=\frac{N(Restrict\cap Attack)}{N({Restrict})},
    Rec=\frac{N(Restrict\cap Attack)}{N({Attack})},
    F1=\frac{2\times Pre\times Rec}{Pre+Rec}
\end{equation}
where $N(\cdot)$ calculates the number of sentences; $Restrict$ indicates the sentences restricted by LLMs, and $Attack$ indicates the ground-truth sentences that contain the attack methods and relations.

\begin{table}[htbp]
  \centering
  \small
  \caption{The performances of various LLMs on restricting sentences that cover the attack trees. (\%).}
      \vspace{-0.2cm}
    {
    \begin{tabular}{l|l|c|c|ccc}
    \toprule
    \multicolumn{1}{c|}{\multirow{2}[0]{*}{\textbf{Method}}} & \multicolumn{1}{c|}{\multirow{2}[0]{*}{\textbf{Model Version}}} & \multicolumn{1}{c|}{\multirow{2}[0]{*}{\textbf{Parameters}}} & \multirow{2}[0]{*}{\textbf{Training Strategy}} & \multicolumn{3}{c}{\textbf{Metrics}}\\
    & & & & Pre. & Rec. & F1\\
    \midrule
    Albert & Albert-large & 18M & Fix-prompt Tuning & 80.93 & 84.65 & 82.75 \\
    T5 & T5-base & 220M & Fix-prompt Tuning & 90.10 & 91.74 & 90.91 \\
    GPT-3 & Text-davinci-003 & 175B & Fix-prompt Tuning & \textbf{97.24} & \textbf{99.36} & \textbf{98.29}\\
    ChatGPT & GPT-3.5-turbo & Unknown & ICL (5-shot) & 76.35 & 88.54 & 82.00 \\
    \bottomrule
        \end{tabular}%
    }
  \label{tab:llm_analysis}%
\end{table}%

Table \ref{tab:llm_analysis} shows the results of different LLMs.
We can see that, the fine-tuned GPT-3 outperforms other models in restricting the scope of sentences with the 97.24\% Precision and 99.36\% Recall.
Therefore, we believe that using GPT-3 with the fix-prompt tuning can accurately restrict the scope of sentences that may contain the attacks.

\subsection{Extracting the Attack Events and Relations}

{In this section, we propose a transition-based model to jointly extract attack events and their relations.
We first define the state, which is a key concept in the transition-based model.
We also define 12 actions that transit the states and extract the events and relations simultaneously while updating the intermediate results.} Finally, we describe the four key steps involved in the transition-based model to extract events and relations.

\subsubsection{The Logic Flow of the Transition-based Event \& Relation Extraction}
\begin{figure*}[t]
\centering
\includegraphics[width=\textwidth]{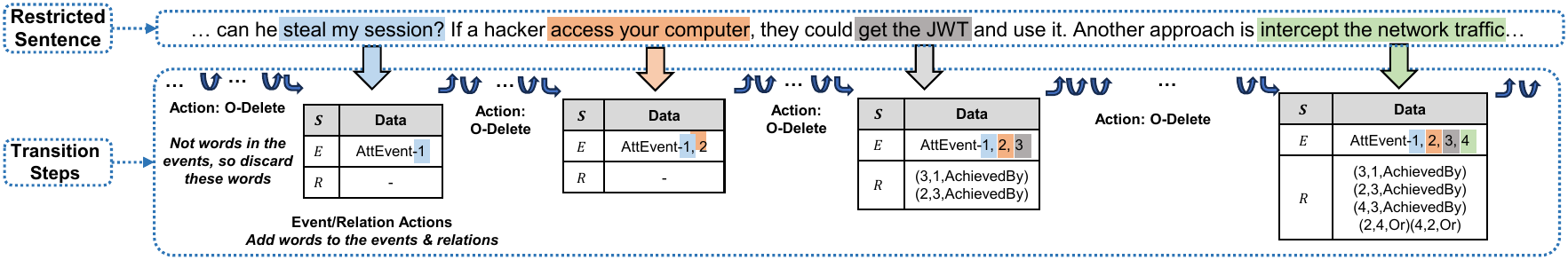}
\caption{{The example of how the transition-based model performs to extract the events and relations from the restricted scope of sentences in motivation example (Fig. \ref{fig:motivation}).}}
\label{fig:case_for_transition}
\end{figure*}

{In Section \ref{sec:transtion_motivation}, we have discussed the benefits of the transition-based framework that is utilized in the information extraction tasks, such as joint entity-relation extraction.
However, the structures of the attack events and their relations are more complex than the normal entities, and previous transition-based frameworks cannot be directly applied to our tasks on synthesizing the attack trees.
To extract the attack events and relations, we propose a new transition-based model based on their structures.

The transition-based model needs to traverse all the words in the restricted sentence scope \textbf{once} and preserve all the essential information while traversing the sentence.
When the model traverses to a certain word $w_i$, we utilize the \textbf{State} $\mathcal{S}$ to save the current situation based on the historical words $(w_1,...,w_i)\rightarrow\mathcal{S}_i$, which illustrates which attack events and relations are extracted, and some intermediate information that can be used to predict the next events and relations.
After storing the current state, the model will traverse to the next word $w_{i+1}$, and determine whether the state will be \textbf{updated} based on the newly-traversed word with the \textbf{Action} $\mathcal{A}$.
The action is determined by the current state and the newly-traversed words $(\mathcal{S}_i,w_{i+1})\rightarrow \mathcal{A}_i$.
As illustrated in Fig. \ref{fig:case_for_transition}, we utilize a specific example to illustrate how the transition-based framework performs to extract the events and relations in the restricted scope of sentences, where restricted sentences come from the motivation example in Fig. \ref{fig:motivation}.
If the transition-based model encounters words that do not belong to the events and relations, it will delete them with the action O-DELETE. Otherwise, the model will preserve these words and restore them to the current states, where we separately store the events and relations to the event set ($E$) and relation set ($R$).
From this example, we can intuitively find out that the transition-based model only requires traversing all words once to extract the attack events and relationships, which has a small time and space complexity ($\mathcal{O}(n)$).

In practice, it is quite a challenge to utilize the transition-based model to extract the attack events and relations, mainly due to the following reasons. 
(1) First, the current definition of the state $\mathcal{S}$ only incorporates the event and relation sets, but we also need to design a specific data structure to distinguish the \textbf{Trigger}, \textbf{Arg-Instrument}, and \textbf{Arg-Target} in the attack events, as well as storing the relations between attack events. 
(2) Second, the processed words in each state need to be considered to predict the next action, where some words deleted or inserted into the previous two sets should be involved in the action prediction. To address this, we include the \textbf{stack} in the state to hold the processed and unprocessed words.
(3) Third, the transition-based model requires the appropriate function to map current states to the required actions, and we introduce the StackLSTM to embed the current state, and predict the action with a linear multi-label classifier.}

\subsubsection{The Definition of State}

\begin{table*}[t]
  \centering
  \caption{The formula expression and description of 12 state actions.}
        \vspace{-0.2cm}
    \resizebox{\textwidth}{!}{\begin{tabular}{c|c|c|l|m{8cm}}
    \toprule
    {\textbf{Category}} & {\textbf{Id}} & {\textbf{Action}} & \multicolumn{1}{c|}{\textbf{Formula Expression$^{*}$}} & \multicolumn{1}{c}{\textbf{Description}} \\
    \midrule
    \multirow{13}[0]{*}{\makecell[c]{\textbf{Event}\\\textbf{Actions}}}& $\mathcal{A}_1$ & {GEN-TRG} & 
    {${\{[j|e]\}\Rightarrow\{e, [j^{t}|\beta], E\cup\{j^{t}\}\}}$} & The word $j$ is predicted as trigger in $e$, so it is marked as $j^{t}$ and moved to $\beta$, and $j^{t}$ is inserted into set $E$. \\
\arrayrulecolor{lightgray}\cline{2-5}
          & $\mathcal{A}_2$ &{GEN-ARG} & {${\{[j|e]\}\Rightarrow\{e, [j^{a}|\beta], E\cup\{j^{a}\}\}}$} & The word $j$ is predicted as argument. so it is marked as $j^{a}$ and moved to $\beta$, and $j^{a}$ is inserted into set $E$. \\
          \cline{2-5}& $\mathcal{A}_3$
          & {DEP-SHIFT} & {${\{[\xi|i^{t}], [j^{a}|\beta]\}\Rightarrow\{[\xi|i^{t}], [\tau|j^{a}], \beta, E\cup\{i^{t}\leftrightarrow j^{a}\}\}}$} & Construct the one-to-one dependency with the top argument $j^{a}$ in $\beta$ and the top trigger $i^{t}$ in $\xi$, insert it into set $E$, and push $j^{a}$ back to the $\tau$. \\
          \cline{2-5}& $\mathcal{A}_4$
          & {TRG-PASS} & {${\{[\xi|i^{t}], [j^{a}|\beta]\}\Rightarrow\{\xi, [i^{t}|\xi '], E\cup\{i^{t}\leftrightarrow j^{a}\}\}}$} & Construct the one-to-many dependency with the top argument $j^{a}$ in $\beta$ and the top trigger $i^{t}$ in $\xi '$, and insert it into event state $E$. \\
          \cline{2-5}& $\mathcal{A}_5$
          & {ARG-PASS} & {${\{[\tau|i^{a}], [j^{t}|\beta]\}\Rightarrow\{\tau,[i^{a}|\tau '], E\cup\{i^{a}\leftrightarrow j^{t}\}\}}$} & Construct the one-to-many dependency $D_{ij}$ with the top trigger $j$ in $\beta$ and the top argument $i$ in $\tau '$, and insert it into event state $E$. \\
    \arrayrulecolor{black}\hline
    \multirow{10}[0]{*}{\makecell[c]{\textbf{Relation}\\\textbf{Actions}}}& $\mathcal{A}_6$ & {RIGHT-SHIFT} & {${\{[\xi|i^{t}], [j^{t}|\beta]\}\Rightarrow\{[\xi|i^{t}, j^{t}], \beta, R\cup\{j^{t}\rightarrow i^{t}\}\}}$} & Construct the one-to-one right-to-left relation from the top trigger $i^{t}$ in $\xi$ to the top trigger $j^{t}$ in $\beta$, insert it to the set $R$, and push the $j$ to the $\xi$.\\
    \arrayrulecolor{lightgray}\cline{2-5}& $\mathcal{A}_7$
          & {RIGHT-PASS} & ${\{[\xi|i^{t}], [j^{t}|\beta]\}\Rightarrow\{\xi, [i^{t}|\xi '], R\cup\{j^{t}\rightarrow i^{t}\}\}}$ & Construct the one-to-many right-to-left relation. Shift the $i^{t}$ from $\xi$ to $\xi '$, construct a relation from $i^{t}$ to the top trigger $j^{t}$ in $\beta$, insert it to the set $R$. \\
          \cline{2-5}& $\mathcal{A}_8$
          & {LEFT-SHIFT} & {${\{[\xi|i^{t}], [j^{t}|\beta]\}\Rightarrow\{[\xi|i^{t}, j^{t}], \beta, R\cup\{j^{t}\leftarrow i^{t}\}\}}$} & Construct the one-to-one left-to-right relation from the top trigger $j^{t}$ in $\beta$ to the top trigger $i^{t}$ in $\xi$, insert it to the set $R$, and push the $j^{t}$ to the $\xi$. \\
          \cline{2-5}& $\mathcal{A}_9$
          & {LEFT-PASS} & {${\{[\xi|i^{t}], [j^{t}|\beta]\}\Rightarrow\{\xi, [i^{t}|\xi '], R\cup\{j^{t}\leftarrow i^{t}\}\}}$} & Construct the one-to-many left-to-right relation. Shift the $i^{t}$ from $\xi$ to $\xi '$, construct a relation from the top trigger $j^{t}$ in $\beta$ to $i^{t}$, insert it to the set $R$.\\
    \arrayrulecolor{black}\hline
    \multirow{4}[0]{*}{\makecell[c]{\textbf{Word}\\\textbf{Actions}}}& $\mathcal{A}_{10}$ & {GEN-SHIFT} & {${\{[j|\beta]\}\Rightarrow\{[j|e]\}}$} & Process a word $j$. Shift a word $j$ from stack $\beta$ to $e$ for further processing. \\
    \arrayrulecolor{lightgray}\cline{2-5}& $\mathcal{A}_{11}$
    & {NO-SHIFT} & {${\{[i^{t}|\xi '],[j^{a}|\tau '], [k^{t},l^{a}|e]\}\Rightarrow\{[\xi|i^{t},k^{t}], [\tau|j^{a},l^{a}]\}}$} & Store the processed word. Push the triggers and arguments from $\xi '$, $\tau '$, $e$ to the $\xi$ and $\tau$. \\
          \cline{2-5}& $\mathcal{A}_{12}$
    & {O-DELETE} & {${\{[j|\beta]\}\Rightarrow\{\beta\}}$} & Delete a word $j$ from state $\beta$. \\
    \arrayrulecolor{black}\bottomrule
    \end{tabular}}%
    \begin{tablenotes}
    \footnotesize
    \item[*] * \textbf{Note}: The symbol $[X^Y|Z]$ indicates that the word $X$ has the type $Y$ (trigger $t$ or argument $a$), and it is pushed into the top of Stack $Z$. 
    \end{tablenotes}
  \label{tab:actions}%
\end{table*}%

To accommodate the transition-based joint event and relation extraction, we first define a {transition} state $\mathcal{S}=\{\xi,\tau,\xi ',\tau ',e,\beta, E, R\}$ to represent the key information regarding attack events and relations,
where the eight elements of $\mathcal{S}$ are described as follows:

\begin{itemize}[leftmargin=*]
    \item $\xi$ and $\tau$ are stacks to store the words of processed triggers and arguments, respectively.
    \item $\xi '$ and $\tau '$ are stacks to store the words that were popped out of $\xi$ and $\tau$ but will be pushed back later.
    \item $e$ is the stack holding the words of partial trigger and action.
    \item $\beta$ is a buffer holding the unprocessed words in the candidate sentence.
    \item $E$ and $R$ are two sets, which respectively store the predicted events (triggers, arguments, and their dependencies), and relations between events predicted during the extraction process.
\end{itemize}

{Based on the event set $E$, we can formulate an event $E_i$ with a trigger $i^{t}$, and arguments $j^{a}$, stored in the set that depend on the trigger. That is, $E_i=(i^{t},j^{a},arg\_type)$, where $arg\_type$ is the type of argument, i.e., Target and Instrument.
The relation between two events $E_i$ and $E_j$ is formulated as $R_{ij}=(E_i,E_j,rel\_type)$, where $rel\_type$ is the type of relation, i.e., {\andrelation}, {\orrelation} and {\parentrelation}.}

\subsubsection{The Definition of Action}

We define 12 actions $\{\mathcal{A}_1,\mathcal{A}_2,...,\mathcal{A}_{12}\}$ to transit the states, and reason attack events and their relations during the extraction process. As shown in Table~\ref{tab:actions}, these actions are in three categories:

\begin{itemize}[leftmargin=*]
    \item {\textbf{Event Actions ($\mathcal{A}_1\sim \mathcal{A}_5$)}}, which are utilized to recognize triggers {(GEN-TRG, TRG-PASS)} and arguments {(GEN-ARG, ARG-PASS)} of attack events, and the dependencies between them {(DEP-SHIFT)};
    \item {\textbf{Relation Actions ($\mathcal{A}_6\sim \mathcal{A}_9$)}}, which are used to recognize relations between event pairs (RIGHT/LEFT-SHIFT, RIGHT/LEFT-PASS);
    \item {\textbf{Word Actions ($\mathcal{A}_{10}\sim \mathcal{A}_{12}$)}}, 
    which are used to process (GEN-SHIFT) and delete (O-DELETE) an unprocessed word, or store the processed words (NO-SHIFT).
\end{itemize}

\subsubsection{Event and Relation Extraction}

\begin{algorithm}[t]
\small
	\caption{Process of transition-based attack events and relations extraction.} 
 \label{alg:transition}
	\KwIn{The words in restricted sentence $S_{atk}$: $\{w_1,w_2...,w_m\}$; transition state $\mathcal{S}$; and transition actions $\{\mathcal{A}_1,\mathcal{A}_2,...,\mathcal{A}_{12}\}$.} 
	\KwOut{The terminal state $\mathcal{S}_T$ with the extracted events and relations in $\{E, R\}\in \mathcal{S}_T$.} 
    \tcp{\textbf{Step 1:} Initializing the state to $\mathcal{S}_0$, with the position-based word embeddings.}
    $\{\vec{w}_1,\vec{w}_2,...,\vec{w}_m\}=PosEmbedding\{w_1,w_2...,w_m\}$, $[\{\vec{w}_1,\vec{w}_2,...,\vec{w}_m\}|\beta]$, where $\mathcal{S}=\mathcal{S}_0$\;
    \tcp{\textbf{Step 2:} Predicting the actions and updating the states, which are the iterations of state-action transition.}
    \While{State Stacks $\{\xi ',\tau ',e,\beta\}\in\mathcal{S}$ are empty, where $\mathcal{S}=\mathcal{S}_T$} 
		{ 
			$\vec{\mathcal{S}}=StackLSTM(\mathcal{S})$\;
            $\mathcal{A}_i=\arg\max_{\mathcal{A}_i}{ p(\mathcal{A}_i|\mathcal{S})}=\arg\max_{\mathcal{A}_i}{\sigma_{\mathcal{A}}(\vec{\mathcal{S}})}$, where $\sigma_{\mathcal{A}}$ is the multi-label classifier\;
            $\mathcal{S} \stackrel{\mathcal{A}_i}{\longrightarrow} \mathcal{S}_{\rm next}$, where the details of transition steps are shown in Table \ref{tab:actions}\;
		}
    \For {$\{i^{t}\leftrightarrow j^{a}\}\in E$ and $\{i^{t}\leftrightarrow j^{t}\}\in R$}{
    \tcp{\textbf{Step 3:} Predicting the type of the events and relations, with the sets in $\{E,R\}\in\mathcal{S}_{T}$.}
    $E_{i}.arg\_type=\arg\max_{arg\_type}{p(arg\_type|i^{t},j^{a})}=\arg\max_{arg\_type}{\sigma_{E}(\vec{j^{a}},\vec{i^{t}})}$, where $\sigma_{{E}}$ is the multi-label classifier, and $arg\_type\in\{\text{Target, Instrument}\}$\;
    $\vec{E}_i=\vec{i^{t}}\oplus\{\vec{j_1},\vec{j_2},...,\vec{j_n}\}$, where all the arguments $j$ depend on $i^t$ in $E_i$: $i^{t}\leftrightarrow\{j_1,j_2,...,j_n\}$\;
    $R_{ij}.rel\_type=\arg\max_{rel\_type}{p(rel\_type|E_i,E_j)}=\arg\max_{arg\_type}{\sigma_{R}(\vec{E_i},\vec{E_j})}$, where $\sigma_{{R}}$ is the multi-label classifier, and $rel\_type\in\{\text{{\andrelation}, {\orrelation}, {\parentrelation}}\}$\;
    \tcp{\textbf{Step 4:} Identifying the implicit relations, with \textbf{Transitive Rule} and \textbf{Asymmetric Rule}.}
    \textbf{Transitive:} $(R_{ij}.rel\_type={\parentrelation},R_{jk}.rel\_type\neq\emptyset)\Rightarrow(R_{ik}.rel\_type={\parentrelation})$\;
    \textbf{Asymmetric:} $(R_{ij}.rel\_type={\parentrelation})\Rightarrow(R_{ji}.rel\_type\neq{\parentrelation})$\;
    }
    return $\{E,R\}$;
\end{algorithm}

{Our transition-based event and relation extraction is an iterative process. In each iteration, the model predicts an action based on the current state and then uses the predicted action to update the state, which will be used to predict the action in the next iteration. Such process continues until the model terminates (i.e., stacks $\xi ',\tau ',e,\beta$ become empty) or a bound on the number of iterations is reached. There are four key steps involved in the process, as is shown in Algorithm \ref{alg:transition}:}

{
\textbf{Step 1: Initializing states (Line 1 in Algorithm \ref{alg:transition}).}
Given the restricted sentence $s_{atk}$ (Section~\ref{sec:limiter}), we embed each inside word in a combined way to better capture the contextual information, and initialize the state $\mathcal{S}$ with the word embedding.
We first embed each word with the position-based word embedding (\textit{PosEmbedding})~\cite{transformer}, which is a widely-used method that can embed the words with its position information, and improve the accuracy of joint extraction~\cite{DBLP:conf/acl/ZhengWBHZX17}.
Then, we initialize the unprocessed words stack $\beta\in\mathcal{S}_0$ with the embedded words, and keep the other states empty.

{\textbf{Step 2: Predicting Actions and Updating States (Line 2$\sim$6).}}
After initializing the state $\mathcal{S}_0$, we input it into the transition iterations.
In each iteration, {\tool} predicts an action based on the current state. 
We first represent the state to $\vec{\mathcal{S}}$ with the Stack LSTM \cite{DBLP:conf/acl/DyerBLMS15}, which can embed the state stacks with the push\&pop order of words.
Then, we introduce a multi-label classifier with the fully connected layer $\sigma_{\mathcal{A}}$, which can predict the probability of each of the 12 actions we have defined.
{The predicted action $\mathcal{A}_i$ with the highest probability will update the transition state $\mathcal{S}$, which will further be used to predict the action in the next iteration.}

\textbf{Step 3: Predicting Events and Relations (Line 8$\sim$10).}
After the iteration reaches the terminal, we first predict the argument type with the event set $E\in\mathcal{S}_T$.
To decide whether the argument {type} in event $E_i$'s dependency between trigger $i^{t}$ and argument $j^{a}$ (i.e., $i^{t}\leftrightarrow j^{a}$) is \textbf{Instrument} or \textbf{Target},
we feed the word embeddings of them into another multi-label classifier $\sigma_{E}$ and choose the argument type with the highest probability $E_i.arg\_type$.
After all the argument types of $E_i$ have been decided, we concatenate the embeddings of trigger word $\vec{i^{t}}$ and all the dependent arguments $\{\vec{j}_1,\vec{j}_2,...,\vec{j}_n\}$, and build the event embedding $\vec{E}_i$.

Second, we predict the relation types with the relation set $R\in\mathcal{S}_T$.
To decide whether the relation type between event $E_i$ and $E_j$ (i.e., $R_{ij}$) is \textbf{{\andrelation}}, \textbf{{\orrelation}} or \textbf{{\parentrelation}}, we introduce the third multi-label classifier $\sigma_{R}$, and feed the embeddings of events $\vec{E}_i$ and $\vec{E}_j$ into the classifier. The relation type that has the highest probability is chosen as the type for relation ${R}_{ij}.rel\_type$.

\textbf{Step 4: Identify the Implicit Relations (Line 11$\sim$12).}
After predicting the types of all the events and relations, according to the attack tree structure, defined by Vidhyashree et al.~\cite{7943455}, we propose two rules to identify relations {that are not explicitly predicted}, i.e., \textbf{Transitive Rule} and \textbf{Asymmetric Rule}.
\begin{itemize}[leftmargin=*]
    \item \textbf{Transitive Rule}: The {\parentrelation} relation has transitivity, i.e., if the attack event $E_i$ can be achieved by $E_j$, and $E_j$ has any relation with $E_k$, then $E_i$ can be achieved by the event $E_k$.  
    \item \textbf{Asymmetric Rule}: The attack tree should not contain the circles, i.e., if attack event $E_i$ can be achieved by $E_j$, then $E_j$ cannot be achieved by $E_i$.
\end{itemize}

After these four steps, we can obtain the final events and relations in the two sets $E$ and $R$.}

\subsubsection{Fine-tuning}
{We use gradient descent to fine-tune the three multi-label classifiers in the transition-based event and relation extraction model.
After extracting the events and relations, we utilize the combined loss to jointly optimize the classifiers as follows:}
\begin{equation}
\mathcal{L}=\lambda\cdot\underbrace{SSVM(y_E,y_R,p(arg\_type|E),p(rel\_type|R))}_{\mathcal{L}_{Extract}}+(1-\lambda)\cdot\underbrace{H(y_{\mathcal{A}},p(\mathcal{A}|\mathcal{S}))}_{\mathcal{L}_{\mathcal{A}}}
\end{equation}
where $\mathcal{L}$ is the combined loss, and $\lambda$ is the loss-balancing parameter. {where $y_{E}$, $y_{R}$, and $y_{\mathcal{A}}$ are ground-truth argument-type in set $E$, relation-type in set $R$, and the iterations' actions.}

The combined loss $\mathcal{L}$ consists of two parts, and the first part is the loss for event and relation extraction $\mathcal{L}_{Extract}$. Given the probabilities of the event's argument-type prediction (\textbf{Line 8 in Algorithm \ref{alg:transition}}) and the relation-type prediction (\textbf{Line 10}), we calculate the loss based on the SSVM loss, which is the widely-used loss for joint event and relation extraction~\cite{DBLP:conf/emnlp/HanNP19}.

The second part is the loss for action prediction $\mathcal{L}_{\mathcal{A}}$, where the function $H(y,p_x)$ calculates the cross-entropy loss between the probabilities of action prediction (\textbf{Line 4}), and ground-truth action labels. We iteratively optimize the {\tool} until the combined loss achieves convergence.

\subsection{Synthesizing Attack Trees}\label{sec:rules}

In this step, we aim to synthesize the attack trees with three rules based on the extracted attack events and their relations. 
Given the extracted events and relations, we compare the differences between attack trees and these extracted results, and more details are shown in Section \ref{sec:motivation}. 

{Then, we ask the three security practitioners to help us design three synthesizing rules to synthesize the attack trees according to their structures~\cite{7943455}. 
The same three security practitioners have over five years of experience in software security, and they will also participate in the ground-truth labeling of our dataset in the following section.
We introduce the \textbf{Open Card Sorting}~\cite{cardsorting1} to devise the rules that synthesize the attack trees, which is a flexible classification method that allows us to create information categories freely, thus helping designers develop more appropriate types.
Specifically, the rule designers follow three criteria when analyzing the differences between the extracted event \& relations and the ground-truth attack trees in the training dataset. 

\begin{itemize}[leftmargin=*]
    \item \textbf{Criterion 1 (Node Creation):} We ask the rule designers to independently define the rules that reflect the content of nodes in the synthesized attack trees based on the extracted events.
    \item \textbf{Criterion 2 (Edge Creation):} We ask the rule designers to independently define the rules that reflect the layer between the attack tree's nodes with the extracted relation, and find the attack goals in the tree.
    \item \textbf{Criterion 3 (Node/Edge Deletion):} We ask the rule designers to discuss with each other and remove some rules that cannot properly describe the nodes and edges in the attack trees, then we uniform and simplify the expressions of these rules.
\end{itemize}

We conduct over 10 iterations of open card sorting with the rule designers and analyze over 500 pairs of extracted events \& relations and attack trees, and we also ask HUAWEI's security practitioners to utilize the attack trees in their company's security knowledge base when designing the rules.
Finally, we devise the following three rules to synthesize the attack trees, which can map the events to the attack goals and methods, as well as map the relations to the tree layers and logic gates.}

\begin{itemize}[leftmargin=*]
    \item \textbf{Rule 1 (Identifying Attack Goal):} The attack goal is the final target of all the attack methods. Therefore, the attack event that the attacker can achieve indicates the attack tree's goal. The content pattern for the attack goal node is that ``\textit{\textbf{attackers achieve (i.e., the event's [trigger] that represents how to achieve the target) a certain target (i.e., the event's [target] argument)}}'', and we use the attack event to replace the content. For example, the root attack event is ``Attack Event 1'' in Fig. \ref{fig:eventrelation}, which has a trigger ``\textit{\textbf{Steal}}'' and a target argument ``\textit{\textbf{Session}}''. Thus, the content for the attack goal in the synthesized attack tree is ``\textit{\textbf{Steal session}}'', as shown in Fig. \ref{fig:motivation}(b).

    \item \textbf{Rule 2 (Identifying Attack Methods):} For the other events that are not identified as attack goals, we identify them as multiple attack methods that can be used to achieve the attack goal. The content pattern for each attack method node is that ``\textbf{\textit{attackers use (i.e., the event's [trigger] that represents how to use the instrument) the instrument (the event's [instrument] argument) to attack a certain target (the event's [target] argument}}'', and we use the corresponding attack events to replace them respectively.

    \item \textbf{Rule 3 (Constructing Gates \& Layers):} The {\parentrelation} relation indicates the layers of attack trees, and the {\andrelation} and {\orrelation} relations indicate the type of logic gates between all the attack methods in the same layers. Therefore, we construct the attack trees with the relations. For example, the logic gate between ``Attack Event 2'' and ``Attack Event 4'' in Fig. \ref{fig:eventrelation} is \textit{\textbf{{\orrelation} Gate}}. The attack methods identified from these two methods are the sub-node of the ``Attack Event 3'' since they both have the {\parentrelation} relation with it.
\end{itemize}

Finally, {\tool} synthesizes attack trees from the security discussions with the previous three rules, and we store them as crowd security knowledge for future reuse.

\section{Experimental Design}
{To evaluate the performance of {\tool}, we investigate the following three research questions:
\begin{itemize}[leftmargin=*]
    \item \textbf{RQ1: What is the performance of {\tool} on synthesizing attack trees from the security discussions?}
    \item \textbf{RQ2: What is the performance of {\tool} on extracting events from the security discussions}
    \item \textbf{RQ3: What is the performance of {\tool} on extracting relations from the security discussions?}
    \item \textbf{RQ4: What is the performance of {\tool} on enhancing the security knowledge bases?}
\end{itemize}
}

\subsection{Security Dataset Preparation}
To evaluate the performance of {\tool}, we prepare the security dataset from Stack Overflow's security posts with the following four steps, i.e., security post collection with security-related tags, data preprocessing, ground-truth labeling, and data augmentation.

\textbf{Step 1: Security Post Collection.}
{Following the previous study~\cite{DBLP:journals/corr/abs-2008-04176}, we collect security posts from Stack Overflow via Stack Exchange \cite{StackExchange}.}
{Specifically, we retrieve security posts from the beginning until Jan 1st, 2023, and prepare the dataset in the following steps: 
(1) We retrieve all the posts in Stack Overflow that have security-related tags (i.e., $<$\textit{security}$>$, $<$\textit{websecurity}$>$, and $<$\textit{firebase}$>$, etc.). Among the posts in Stack Overflow, we find that posts with specific tags may contain over 90\% of security-related knowledge, such as the attack and mitigations for the different threats;
(2) We remove the posts that receive negative scores voted by {Stack Overflow users}, as well as non-English posts;
and (3) We select the posts that have an accepted answer, and this answer receives the highest score in the post to reduce the bias in subjective decisions.}

\textbf{Step 2: Data Preprocessing.} 
{We preprocess the collected security posts as follows: (1) First, we remove stopwords, correct any typos, and lemmatize the posts with Spacy~\cite{spacy.io}. 
(2) Then, we replace the code blocks (wrapped by HTML tags $<$code$>$, $</$code$>$), with token \texttt{[CODE]}.
(3) Finally, we remove other HTML tags, such as $<$a$>$, $<$li$>$, etc., and retrain the plain text inside these HTML tags.}

\textbf{Step 3: Ground-truth Labeling.}
For each security post, we label the triggers and arguments based on BIO tagging \cite{DBLP:conf/acl/ZhengWBHZX17}, and label the events and relations with Han's tagging \cite{DBLP:conf/emnlp/HanNP19}.
To guarantee the correctness of the labeling result, we build a {13-people} team with three professors, {two Ph.D. students, five Master's students, and three security practitioners of HUAWEI}. {All of them have rich experience in secure software development, and six out of 13 (46\%) participants are external annotators to reduce the bias of labeling.}
Each security post is labeled by two different team members. 
{When different opinions occur on the labels, we discuss them with all team members until a decision has been reached.}
{Only a few attack trees have conflicting opinions, and the average Cohen’s Kappa \cite{DBLP:journals/jss/PerezDMT20} is 0.87\footnote{The magnitude guidelines define that kappa value above 0.81, which means almost perfect agreement among all the team members.}, meaning the biases of ground-truth labeling are minor.}

\begin{table}[t]
  \centering
  \small
  \caption{
  The statistics of our dataset.
  }
  \vspace{-0.2cm}
    {\begin{tabular}{c|cccc|cccc}
    \toprule
    \multirow{2}[0]{*}{\textbf{Category}} & \multicolumn{4}{c|}{\textbf{Original}} & \multicolumn{4}{c}{\textbf{Augmented}} \\
          & \#post & \#tree & \#event & \#relation & \#post & \#tree & \#event & \#relation \\
    \midrule
    \textit{Train} & 3,042 & 868  & 1,685 & 4,625 & 16,175 & 13,425 & 30,172 & 89,629 \\
    \textit{Valid}     & 1,014 & 189 & 497 & 1,255 & 1,014 & 189 & 497 & 1,255 \\
    \textit{Test}     & 1,014 & 297 & 828 & 1,876 & 1,014  & 297 & 828 & 1,876 \\
    \hline
    \textit{Total} & \textit{5070}  & \textit{1,354} & \textit{3,010} & \textit{7,756} & \textit{18,203} & \textit{13,911} & \textit{31,497} & \textit{92,760} \\
    \bottomrule
    \end{tabular}%
}
\vspace{-0.3cm}
  \label{tab:datatable}%
\end{table}%

\textbf{Step 4: Data Augmentation.} 
For training and testing the {\tool}, we split the dataset into training, validating, and testing datasets with the proportion 60\%, 20\%, and 20\%.
{Since the number of trained posts with attack trees is much lower than posts without attack trees, as shown in \textbf{Original} column of Table \ref{tab:datatable}, we employ two strategies to augment the training dataset.} 
The first strategy is to use bootstrap sampling \cite{bootstrap}, which randomly copies the posts with events and relations. 
The second strategy is EDA~\cite{DBLP:conf/emnlp/WeiZ19} augmentation, which is a typical data augmentation technique. 
While keeping the events and relations not affected by data augmentation, EDA randomly inserts, deletes, and swaps several words in the source security post with a certain probability.
{The results of our data augmentation are shown in the \textbf{Augmented} column}.

Table \ref{tab:datatable} shows the size of the dataset, where the column ``\textbf{Original}'' shows the number of posts, trees, events, and relations in the original dataset, and the column ``\textbf{Augmented}'' shows the size of the expanded dataset. In total, we have collected 5,070 security posts initially and expanded the dataset to 18,203 security posts.


\subsection{Baselines}

Since our work is the first to synthesize attack trees from security posts, there are currently no directly comparable approaches. Therefore, we use the SOTA approaches of the most similar NLP tasks, i.e., \textit{attack tree synthesizing} task (RQ1), \textit{attack event extraction} task (RQ2), and \textit{relation extraction} task (RQ3). To make these approaches fit our task, we fine-tune them with the same settings as our approach, so that they can output the attack trees, attack events, and attack-event relations.
Note that, to fairly compare with baselines, we retrain and test them under the same dataset and parameter settings as {\tool}.

\textbf{Baseline for RQ1.}
In RQ1, we compare the {\tool} with the two types of baselines.
The first type is the joint event and relation extraction model, i.e., the \textbf{Structured-Joint}~\cite{DBLP:conf/emnlp/HanNP19}, which is the SOTA model to extract both event and relation in parallel, with shared representation learning and structured prediction. 
To accommodate Structured-Joint to the attack tree synthesizing, we extend it to \textbf{Structured-Joint+} by (1) retraining with the same labeled event and relation data as our {\tool}; (2) applying our heuristic rules to help Structured-Joint synthesize attack trees.
{The second type is the aforementioned LLM, which is effective in various NLP tasks. We choose three SOTA generative LLMs to synthesize the attack trees, i.e., \textbf{Albert}~\cite{DBLP:conf/iclr/LanCGGSS20}, \textbf{T5}~\cite{DBLP:journals/jmlr/RaffelSRLNMZLL20}, and \textbf{GPT-3}~\cite{DBLP:conf/nips/BrownMRSKDNSSAA20}, by utilizing the follow-up prompt template:
\begin{center}
\vspace{-0.2cm}
\small
\begin{tcolorbox}[colback=white,
                  colframe=black,
                  width=\columnwidth,
                  arc=1mm, auto outer arc,
                  boxrule=0.4pt,
                  left=0.1pt,
                  right=0.1pt,
                  top=0.1pt,
                  bottom=0.1pt,
                colbacktitle=white!80!gray, coltitle=black, 
                title={\textbf{Cloze-Template Prompt for LLM Baselines}}
                 ]
{\textit{\textbf{Cloze-Testing:} From the following security post \{Question, Accepted\_Answer\},  (Title is \{Title\} to summarize the main topic of security post) \textbf{[Y]} is the the synthesized attack tree. (Omit the definition of attack trees)}}

\textit{\textbf{Task:} Please predict the token [Y] with the generated attack trees from the security post's question and answers to make the cloze-testing complete. The format of the attack tree is a dictionary. We use the following example to show the format.}

\textit{\textbf{Output Example:} \{Nodes: [\{$G$: steal session\}, \{$M_1$: get JWT to attack session\}, \{$M_2$: access computer to attack JWT\}, \{$M_3$, intercept network traffic to attack JWT\}], Edges: [($G$, $M_1$, {\parentrelation}), ($M_1$, $M_2$, {\parentrelation}), ($M_2$, $M_3$, {\orrelation})]\}}
\end{tcolorbox}
\end{center}
where the \textit{Question}, \textit{Accepted\_Answer}, and \textit{\{Title\}} are the input sentences and title of the security post, and \textit{[Y]} is the text for representing the output attack trees, which contains a set of tuples~\cite{schneier1999attack}. 
The output of the attack tree is organized as a dictionary \{Nodes: $Node\_List$, Edges: $Edge\_List$\}. 
In the $Node\_List$, $G$ is the attack goal, and $M_i$ indicates the $i_{th}$ attack method.
In the $Edge\_List$, we use the triplets to represent the edges, where the first two elements indicate the nodes concatenated by the edge, and the third element is the relation. 
We also use the output example of Figure \ref{fig:motivation} to guide the LLMs to output the attack trees.



\textbf{Baselines for RQ2.}
{We first compare the {\tool} with the previous joint event and relation extraction model, i.e., \textbf{Structured-Joint}, and the best-performed LLM in RQ1, i.e., \textbf{GPT-3}. 
Then we compare {\tool} with two representative baselines on event extraction, which obtain the SOTA performances on our dataset.}
{\textbf{ED3C} \cite{DBLP:conf/emnlp/VeysehNNMN21} introduces the context selection to reduce the noise and improve the representation of events in a large-scale document; and
\textbf{MLBiNet} \cite{DBLP:conf/acl/LouLDZC20} propagates event-based semantic information across sentences and extract multiple events from the complex documents.}

\textbf{Baselines for RQ3.}
{We first compare the {\tool} with the joint event and relation extraction model, and best-performed LLM in RQ1, i.e., \textbf{Structured-Joint} and \textbf{GPT-3}. 
Then we compare {\tool} with two representative baselines on event extraction, which obtain SOTA performances on our dataset.}
\textbf{KnowledgeILP} \cite{DBLP:conf/emnlp/NingSR19} is a data-driven method which incorporates ILP and commonsense knowledge; and
\textbf{JC Learning} \cite{DBLP:conf/emnlp/WangCZR20} is a joint-constrained learning model for multi-faceted event-event relation extraction.
{In addition, we also compare the \textbf{Time Cost} (i.e., {Training/Fine-tuning Hours}) of {\tool} and baselines. We calculate the hours of training the {\tool} and fine-tuning the baselines on our dataset.}


\subsection{Evaluation Metrics}\label{sec:metrics}

{To evaluate the performance of attack tree synthesizing, we employ two widely used metrics of tree similarity \cite{DBLP:phd/ethos/Lin10c}: {Average Hamming Distance}  and {Tree-editing Distance Similarity}, reflecting to what extent the synthesized attack trees are similar to the ground truth. 
(1) \textbf{Average Hamming Distance} (\textbf{AHD}) \cite{DBLP:journals/actaC/KeriK04} aims to measure the difference between attack goals and methods, which calculates the average proportion of deviation words between nodes at the same position on synthesized and ground-truth attack tree, indicating the node similarity between two trees.
(2) \textbf{Tree-editing Distance Similarity} (\textbf{TEDS}) \cite{DBLP:journals/sigmod/LiWLG13} aims to measure the relation extraction based on the attack tree, which calculates the average proportion of edges and logic gates to be edited from synthesized to the ground-truth attack tree, indicating the structural similarity between two trees.} 
\begin{equation}
    AHD=\frac{\sum Hamming(Node\_Predict,Node\_Truth)}{\max{N(Node\_Predict,Node\_Truth)}},
    TEDS=\frac{N(Edge\_Predict\rightarrow Edge\_Truth)}{\max{N(Edge\_Predict,Edge\_Truth)}}
\end{equation}
where function $N(\cdot)$ calculates the number of nodes, edges, and editing steps from predicted to ground-truth trees. A lower AHD or TEDS value means a higher level of tree similarity.

To evaluate the performance of event and relation extraction, we use three commonly used metrics: 
(1) \textbf{Precision} (\textbf{Pre.}), which calculates the ratio of correct positive predictions to the total positive predictions; (2) \textbf{Recall} (\textbf{Rec.}), which calculates the ratio of correct positive predictions to the ground-truth positive labels; 
and (3) \textbf{F1-measure} (\textbf{F1}), which calculates the harmonic mean of the precision and recall. 
\begin{equation}
    Pre=\frac{N(ER\_Predict\cap ER\_Truth)}{N({ER\_Predict})},
    Rec=\frac{N(ER\_Predict\cap ER\_Truth)}{N(ER\_Truth)},
    F1=\frac{2\times Pre\times Rec}{Pre+Rec}
\end{equation}
where function $N(\cdot)$ calculates the number of events and relations. The higher of these three metrics indicates the higher performances of event and relation extraction.




\subsection{Experiment Setting}\label{sec:setting}

For {\tool} training on all experiments, we utilize the gird search~\cite{DBLP:journals/ijrr/LaValleBL04} technique to tune the hyper-parameters until the model achieves the highest performances.
Among these hyper-parameters, we set the $batch\_size=8$ for training the {\tool}, and $\lambda=0.5$ for loss balancing.
{\tool} and baselines run on a Windows 10 PC, with NVIDIA GeForce RTX 2060 GPU and 32GB RAM.

For RQ1-3, we aim to first compare the {\tool} with the traditional baselines on attack tree synthesizing. Then, we need to compare it with the LLMs and analyze the advantages of {\tool} compared to directly using the LLMs to synthesize the attack trees, extract attack events, and extract their relations.
We choose the same \textit{batch\_size} of {\tool} to all the baselines, and other parameters are determined by greedy strategy, which can achieve the best performance after tuning these baselines and our approach.

For RQ4, we aim to analyze the ability of synthesized attack trees from security posts and Github issues, to enhance the typical public security bases, i.e., CAPEC~\cite{CAPEC1} and CVE~\cite{CVE}, and the private security bases, i.e., HUAWEI's attack tree database.

\section{Results}\label{sec:results}
\subsection{Performance on attack tree synthesizing}\label{sec:human}

{Table \ref{tab:attack_tree_syn} illustrates the comparison between {\tool} and the four baselines on attack tree synthesizing, and the best performance of each column is highlighted in \textbf{bold face}.}

\begin{table}[htbp]
  \centering
  \small
  \caption{The comparison between {\tool} and baselines on synthesizing the attack trees (\%).}
      \vspace{-0.2cm}
    {
    \begin{tabular}{l|l|c|c|cc}
    \toprule
    \multicolumn{1}{c|}{\multirow{2}[0]{*}{\textbf{Method}}} & \multicolumn{1}{c|}{\multirow{2}[0]{*}{\textbf{Model Version}}} & \multicolumn{1}{c|}{\multirow{2}[0]{*}{\textbf{Parameters}}} & \multirow{2}[0]{*}{\textbf{Time Cost}} & \multicolumn{2}{c}{\textbf{Metrics}}\\
    & & & & AHD & TEDS\\
    \midrule
    Structured-Joint+ & Structured-Joint+ & <10M &  15h & 30.75 & 25.15  \\
    Albert & Albert-large & 18M & 22.7h & 33.16 & 23.74 \\
    T5 & T5-base & 220M & 25h & 26.17 & 20.67 \\
    GPT-3 & text-davinci-003 & 175B & 31.5h & 18.44 & 15.06 \\
        \hline
    \multirow{2}{*}{{\tool}}  & \multirow{2}{*}{{\tool}} & \multirow{2}{*}{<10M (w/o GPT-3)} & \multirow{2}{*}{18h}  & \textbf{10.24} & \textbf{7.93}  \\
    & & & & (\textcolor{red}{$\downarrow$}8.20) &  (\textcolor{red}{$\downarrow$}7.13) \\
    \bottomrule
        \end{tabular}%
    }
  \label{tab:attack_tree_syn}%
\end{table}%

\textbf{Comparison with Baseline Attack Tree Synthesizers.}
{\tool} achieves the lowest both AHD (15.03\%) and TEDS (13.24\%) scores on the performance of synthesizing attack trees, reducing all the baselines by 8.20\% (AHD) and 7.13\% (TEDS). These results indicate that the attack trees generated by {\tool} are much more similar to the ground truth than the baselines.

\textbf{Training/Fine-tuning Hours.}
The training hours of {\tool} are 18h, which is only longer than the fine-tuning hours of Structured-Joint+. Taking both the comparison results and time efficiency into consideration, we believe that {\tool} has advantages over baselines on synthesizing the attack trees.

\begin{figure}[t]
\centering
\includegraphics[width=0.8\columnwidth]{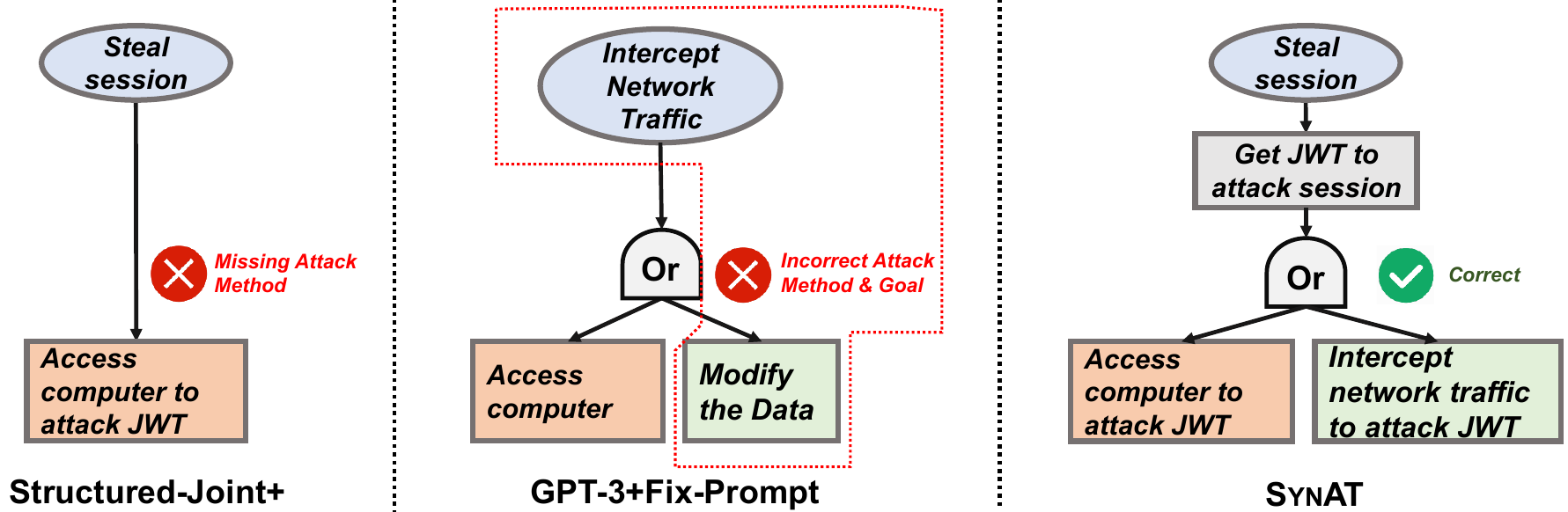}
\caption{The case study of {\tool} on Fig. \ref{fig:motivation}'s example.}
\label{fig:case}
\end{figure}




\textbf{Case Study.}
To qualitatively illustrate the performance of {\tool}, we show the comparison with the example in Fig. \ref{fig:case}. 
We can see that, {\tool} can accurately synthesize the nodes and edges of the attack tree. On the contrary, Structured-Joint+ misses the attack method ``intercept network traffic to attack JWT''; GPT-3 incorrectly predicts the attack goal to ``intercept network traffic'' and predicts an incorrect attack method ``modify the data''.
Therefore, our method outperforms other baselines in attack tree synthesizing of this case.


\textbf{Benefits.}
{We believe the advantages of {\tool} mainly come from the accurate prediction of events and relations from the lengthy post. Compared with the baselines, {\tool} can obtain the events in different sentences and accurately analyze the relation between them, so we can accurately transform them to attack trees with some simple tree synthesizing rules.
}

\begin{center}
\vspace{-0.2cm}
\begin{tcolorbox}[colback=gray!10,
                  colframe=black,
                  width=\columnwidth,
                  arc=1mm, auto outer arc,
                  boxrule=0.4pt,
                  left=0.1pt,
                  right=0.1pt,
                  top=0.1pt,
                  bottom=0.1pt
                 ]
\textit{\textbf{Answering RQ1}: 
{\tool} outperforms the SOTA baseline in synthesizing the attack trees. It reaches the lowest AHD and TEDS at 10.24\% and 7.93\%, indicating that the attack trees generated by {\tool} are much more similar to the ground truth. 
The qualitative analysis further illustrates the effectiveness of {\tool} on the attack tree synthesizing.
} 
\end{tcolorbox}
\end{center}

\subsection{Performance on Event Extraction}
Table \ref{tab:event-extraction} illustrates the comparison results of {\tool} between six baselines on trigger, instrument, and target detection. The best performance of each column is highlighted in \textbf{bold face}. 

\begin{table*}[htbp]
  \centering
  \small
  \caption{The baseline comparison on event extraction (\%).}
  \vspace{-0.2cm}
    \resizebox{\textwidth}{!}{
    \begin{tabular}{l|c|ccc|ccc|ccc|ccc}
    \toprule
    \multicolumn{1}{c|}{\multirow{2}[0]{*}{\textbf{Method}}} & \multicolumn{1}{c|}{\multirow{2}[0]{*}{\textbf{\makecell{Time \\ Cost}}}} & \multicolumn{3}{c}{\textbf{Trigger}} & \multicolumn{3}{|c}{\textbf{Instrument}} & \multicolumn{3}{|c}{\textbf{Target}} &  \multicolumn{3}{|c}{\textit{\textbf{Average}}} \\
          & & {Pre.} & {Rec.} & {F1} & Pre. & Rec. & F1 & Pre. & Rec. & F1 & Pre. & Rec. & F1 \\
    \midrule
    ED3C & 7.7h & 64.17 & 65.35 & 64.75  & 65.46 & 60.45 & 62.86  & 58.76 & 66.84 & 62.54  & 62.80  & 64.21  & 63.38   \\
    MLBiNet & 17h & 71.45 & 67.12 & 69.22  & 70.34 & 69.45 & 69.89  & 67.32 & 75.05 & 70.98  & 69.70  & 70.54  & 70.03   \\
    Structured-Joint+ & 14.5h & 70.35 & 74.24 & 72.24  & 64.27 & 69.08 & 66.59  & 63.24 & 78.03 & 75.86  & 65.95  & 73.78  & 71.56   \\
    GPT-3 & 27h & 72.44 & 73.15 & 72.79 & 70.15 & 71.26 & 70.70 & 70.75 & 77.16 & 73.82 & 71.11 & 73.86 & 72.44  \\
        \hline
    \multirow{2}{*}{{\tool}}  & \multirow{2}{*}{16h} & \textbf{83.36} & \textbf{81.25} & \textbf{82.29} & \textbf{81.93} & \textbf{79.25} & \textbf{80.57} & \textbf{77.83} & \textbf{82.16} & \textbf{79.94} & \textbf{81.04} & \textbf{80.89} & \textbf{80.93}  \\
    && (\textcolor{red}{$\uparrow$}10.92)  & (\textcolor{red}{$\uparrow$}7.01)  & (\textcolor{red}{$\uparrow$}9.50)  & (\textcolor{red}{$\uparrow$}11.78) & (\textcolor{red}{$\uparrow$}7.99)  & (\textcolor{red}{$\uparrow$}9.87)  & (\textcolor{red}{$\uparrow$}7.08)  & (\textcolor{red}{$\uparrow$}4.13)  & (\textcolor{red}{$\uparrow$}4.08)  & (\textcolor{red}{$\uparrow$}9.93)  & (\textcolor{red}{$\uparrow$}7.03)  & (\textcolor{red}{$\uparrow$}8.50) \\
    \bottomrule
        \end{tabular}%
    }
  \label{tab:event-extraction}%
\end{table*}%

\textbf{Comparison with Baselines.}
{\tool} outperforms all the baselines on average,
improving the best baseline in Precision (+9.03\%), Recall (+7.03\%), and F1 (+8.50\%). 
For each trigger and argument detection, {\tool} also reaches the best performance on Precision, Recall, and F1, {with all F1 over 75\%.}
Results indicate that our model is more appropriate to extract events in security posts.

\textbf{Training/Fine-tuning Hours.}
The training hours of {\tool} are 16h, which is higher than the simplest event extraction baseline, i.e., ED3C, and the Structured-Joint. Combining both the comparison results and time efficiency, we believe that {\tool} has the advantage of extracting the attack events.

\textbf{Benefits.}
We believe the advantages come from two aspects: 
(1) {\tool} can extract events from attack sentences instead of the complete security post, which can reduce the impact of noise sentences without attack events. 
(2) {\tool} can combine the information of relations in event extraction, and optimize the event extraction with multiple losses. In this way, {\tool} can adjust the incorrectly extracted events according to relations, which improves its performance on attack event extraction.

\begin{center}
\vspace{-0.1cm}
\begin{tcolorbox}[colback=gray!10,
                  colframe=black,
                  width=\columnwidth,
                  arc=1mm, auto outer arc,
                  boxrule=0.4pt,
                  left=0.1pt,
                  right=0.1pt,
                  top=0.1pt,
                  bottom=0.1pt
                 ]
\textit{\textbf{Answering RQ2}: {\tool} outperforms the four baselines in extracting attack events across the trigger and argument detection, and the average Precision, Recall, and F1 are 81.04\%, 80.89\%, and 80.93\%, outperforming the best baseline by 9.93\% (Pre.), 7.03\% (Rec.), and 8.50\% (F1).}
\end{tcolorbox}
\end{center}

\subsection{Performance on Relation Extraction}

Table \ref{tab:relation-extraction} illustrates the comparison results of {\tool} between baselines on {\parentrelation}, {\andrelation}, and {\orrelation} relation extraction. The best performance of each column is also highlighted in \textbf{bold face}. 

\begin{table*}[htbp]
  \centering
  \small
  \caption{The baseline comparison on relation extraction (\%).}
    \vspace{-0.2cm}

    \resizebox{\textwidth}{!}{
    \begin{tabular}{l|c|ccc|ccc|ccc|ccc}
    \toprule
    \multicolumn{1}{c|}{\multirow{2}[0]{*}{\textbf{Method}}} & \multirow{2}[0]{*}{\textbf{\makecell{Time \\ Cost}}} & \multicolumn{3}{c}{\textbf{{\parentrelation}}} & \multicolumn{3}{|c}{\textbf{{\andrelation}}} & \multicolumn{3}{|c}{\textbf{{\orrelation}}} &  \multicolumn{3}{|c}{\textit{\textbf{Average}}} \\
          & & {Pre.} & {Rec.} & {F1} & Pre. & Rec. & F1 & Pre. & Rec. & F1 & Pre. & Rec. & F1 \\
    \midrule
    KnowledgeILP & 8.5h & 62.06 & 60.45 & 61.24  & 70.33 & 63.44 & 66.71  & 65.87 & 72.13 & 68.86  & 66.09  & 65.34  & 65.60  \\
    JC Learning & 16h & 68.44 & 63.17 & 65.70  & 71.45 & 62.60  & 66.73  & 71.42 & 70.59 & 71.00  & 70.44  & 65.45  & 67.81  \\
    Structured-Joint+ & 14.5h & 75.94 & 62.10  & 68.33  & 73.32 & 68.75 & 70.96  & {74.66} & 81.72 & 78.03  & 74.67  & 70.86  & 72.44 \\
    GPT-3 & 27h &78.22 & 70.15 & 73.97 & 69.06 & 70.15 & 69.60 & 76.24 & 82.16 & 79.09 & 74.51 & 74.15 & 74.22 \\
        \hline

    \multirow{2}{*}{{\tool}}  & \multirow{2}{*}{16h} & \textbf{88.04} & \textbf{89.17} & \textbf{88.60} & \textbf{90.19} & \textbf{93.08} & \textbf{91.61} & \textbf{80.04} & \textbf{86.67} & \textbf{83.22} & \textbf{86.09} & \textbf{89.64} & \textbf{87.81}  \\
    && (\textcolor{red}{$\uparrow$}9.82)  & (\textcolor{red}{$\uparrow$}19.02) & (\textcolor{red}{$\uparrow$}14.64) & (\textcolor{red}{$\uparrow$}16.87) & (\textcolor{red}{$\uparrow$}22.93) & (\textcolor{red}{$\uparrow$}20.65) & (\textcolor{red}{$\uparrow$}3.80)  & (\textcolor{red}{$\uparrow$}4.51)  & (\textcolor{red}{$\uparrow$}4.13)  & (\textcolor{red}{$\uparrow$}11.42) & (\textcolor{red}{$\uparrow$}15.49) & (\textcolor{red}{$\uparrow$}13.59)\\
    \bottomrule
    \end{tabular}%
    }
  \label{tab:relation-extraction}%
\end{table*}%

\textbf{Comparison with Baselines.}
{\tool} outperforms all the baselines on average,
improving the best baseline in Precision (+11.42\%), Recall (+15.49\%), and F1 (+13.59\%).
For each relation, {\tool} also achieves the highest Precision, Recall, and F1, with all F1 over 80\%.
These results indicate that {\tool} is more appropriate for extracting the relations in security posts.

\textbf{Training/Fine-tuning Hours.}
The training hours of {\tool} are 16h, which is higher than the simplest relation extraction baseline, i.e., KnowledgeILP, and the Structured-Joint+. Combining both the comparison results and time efficiency, we believe that {\tool} has the advantage of extracting the relations between attack events.

\textbf{Benefits.}
Apart from the benefits in RQ2, we believe the advantages of relation extraction also come from the constraint loss. 
For example, if the relation ${R}_{ij}$ is predicted to be {\parentrelation}, then {\tool} will adjust ${R}_{ji}$ to {\andrelation} and {\orrelation} instead of {\parentrelation}.
Therefore, constraint loss reduces the probability of rings in the attack trees, thus improving the accuracy of relation extraction.

\begin{center}
\vspace{-0.2cm}
\begin{tcolorbox}[colback=gray!10,
                  colframe=black,
                  width=\columnwidth,
                  arc=1mm, auto outer arc,
                  boxrule=0.4pt,
                  left=0.1pt,
                  right=0.1pt,
                  top=0.1pt,
                  bottom=0.1pt
                 ]
\textit{\textbf{Answering RQ3}: {\tool} outperforms the four baselines in extracting {\andrelation}, {\orrelation} and {\parentrelation} relations, and the average Precision, Recall, F1 are 86.09\%, 89.64\%, 87.81\%, outperforming the best baseline by 11.42\% (Pre.), 15.49\% (Rec.), 13.59\% (F1).}
\end{tcolorbox}
\end{center}

\subsection{Performance on Enhancing Security Knowledge Base}\label{sec:practical}
In this section, we conduct the practical application of enhancing the public security knowledge base, as well as enhancing and private knowledge base (HUAWEI's attack tree database) with its experienced practitioners.

After training the {\tool}, we first collect 192 security posts with the same StackExchange tool in the previous experiments, dated after Feb 1st, 2023 (\textit{note that these posts are \textbf{never used }in the model training or testing}). 
Then, we synthesized 121 new attack trees with the trained {\tool}, and each of them is never modified after they are synthesized.
Third, we invite three security practitioners in HUAWEI to help identify whether these new attack trees can be used to enhance the security knowledge bases. All the practitioners have more than 10 years of experience in maintaining software security of open-source projects, and they mainly care about the attacks and defenses in Web Applications and MongoDB Database. 
We contact the security practitioners via internal emails by sending the attack trees to their email addresses and receiving feedback. Besides, we have no communication with these practitioners, so the influences of subjective evaluation can be alleviated

\textbf{Enhancing Public Security Knowledge base.}
{Based on the 121 synthesized attack trees we synthesized with {\tool}, we ask the security practitioner to analyze their creation times, i.e., whether the synthesized attack trees are proposed earlier in Stack Overflow than that in the public security knowledge base, i.e., CVE and CAPEC.
Although these knowledge bases are comprehensive, they require experienced practitioners to manually analyze the attacks, which may have lags in official information releases.
Moreover, if the attacks have not occurred in these public knowledge bases, we ask the practitioners to analyze whether these attack trees are correct and can be treated as new attacks on the community.

We classify the attacks with IDs in \textbf{Common Weakness Enumeration (CWE)}~\cite{CWE}, which is a widely-used categorization mechanism in these knowledge bases that differentiates the types of vulnerabilities with their causes, behaviors, and consequences.
As a result, the attacks are classified into 10 unique CWE-IDs, from CWE-787 to CWE-125, and most of these CWE-IDs are the Top-20 CWE in 2023.

\begin{table}[htbp]
  \caption{The results of enhancing the public security knowledge base, i.e., CVE and CAPEC.}
  \small
  \vspace{-0.2cm}
\begin{tabular}{l|l|ll|ll}
\toprule
\textbf{\multirow{2}{*}{Category (CWE)}} & \textbf{\multirow{2}{*}{\#{\tool}}} & \multicolumn{2}{c}{\textbf{CVE}} & \multicolumn{2}{|c}{\textbf{CAPEC}} \\
                               &                          & \textbf{\#Early (Ratio)}         & \textbf{\#New (Ratio)}        & \textbf{\#Early (Ratio)}           & \textbf{\#New (Ratio)}        \\
\midrule
CWE-787                                             & 22                                           & 8 (36.36\%)             & 3 (13.64\%)          & 16 (72.73\%)              & 3 (13.64\%)          \\
CWE-79                                              & 17                                           & 6 (35.29\%)             & 4 (23.53\%)          & 12 (70.59\%)              & 5 (29.41\%)          \\
CWE-78                                              & 15                                           & 7 (46.67\%)             & 2 (13.33\%)          & 7 (46.67\%)               & 4 (26.67\%)          \\
CWE-352                                             & 16                                           & 6 (37.50\%)             & 3 (18.75\%)          & 7 (43.75\%)               & 3 (18.75\%)          \\
CWE-190                                             & 14                                           & 7 (50.00\%)             & 3 (21.43\%)          & 7 (50.00\%)               & 3 (21.43\%)          \\
CWE-287                                             & 11                                           & 3 (27.27\%)             & 1 (9.09\%)           & 4 (36.36\%)               & 2 (18.18\%)          \\
CWE-121                                             & 8                                            & 5 (62.50\%)             & 1 (12.50\%)          & 4 (50.00\%)               & 1 (12.50\%)          \\
CWE-138                                             & 6                                            & 2 (33.33\%)             & 1 (16.67\%)          & 3 (50.00\%)               & 1 (16.67\%)          \\
CWE-183                                             & 7                                            & 2 (28.57\%)             & 2 (28.57\%)          & 2 (28.57\%)               & 2 (28.57\%)          \\
CWE-125                                             & 5                                            & 3 (60.00\%)             & 0 (0.00\%)           & 3 (60.00\%)               & 0 (0.00\%)           \\
\hline
\rowcolor{gray!20} All                                                 & 121                                          & 49 (40.50\%)            & 20 (16.53\%)         & 65 (53.72\%)              & 24 (19.83\%) \\
\bottomrule
\end{tabular}
  \label{tab:application_attck}

\end{table}

Table \ref{tab:application_attck} shows the results of enhancing the public security knowledge bases CVE and CAPEC with the attack trees synthesized by {\tool}. 
The column \textbf{\#{\tool}} shows the number of attack trees in each CWE-ID, 
column \textbf{\#Early} indicates the number and ratio of attacks that are proposed in Stack Overflow earlier than the disclosure time, 
and column \textbf{\#New} are the number and ratio of attacks that are potentially new to the knowledge base.
We can see that 40.50\% of the attack trees are created earlier than that in the CVE and 53.72\% are earlier than CAPEC. Also, 16.53\% of the attack trees may contain new attacks that have not been officially reported in CVE, and 19.83\% of them are new to CAPEC.
To sum up, {\tool} can help with enhancing the public security knowledge base.}

\textbf{Enhancing Private Security Knowledge Base.}
Apart from the CVE and CAPEC, we also ask the security practitioners to manually inspect the synthesized 121 attack trees, and finally select trees to replenish their attack knowledge database.
The items in HUAWEI's knowledge base are different attack trees, which include 241 attack trees and carry important information such as attack description, attack case, and possible mitigation in terms of STRIDE~\cite{STRIDE} threat model. 
The database provides developers with code design specifications and threat modeling analysis of {their products}. 
However, the database has not been maintained and updated for a long time, statistically, 77\% of the attack trees have not been updated within four years.
Therefore, they urgently need to use external knowledge to update their security knowledge base.

\begin{table}[t]
  \centering
    \small
  \caption{The result of enhancing HUAWEI's security knowledge base.}
  \vspace{-0.2cm}
    {\begin{tabular}{l|ll|ll}
    \toprule
     \multicolumn{1}{c|}{\textbf{Category}}  & \textbf{\#HWTrees} & \textbf{\#{\tool}} & \textbf{\#Accept (Ratio)} & \textbf{Increment Ratio}\\
    \midrule
    Spoofing                &65     &12     &9 (75.00\%)      &13.85\%\\
    Tampering               &31     &44     &38 (86.36\%)      &122.58\%\\
    Repudiation             &16     &2      &2 (100.00\%)      &12.50\%\\
    Information Disclosure  &54     &41     &34 (82.93\%)      &62.96\%\\
    Denial of Service       &22     &9      &9 (100.00\%)      &40.91\%\\
    Elevation of Privilege  &53     &13     &10 (76.92\%)      &18.87\%\\
    \hline
    \rowcolor{gray!20}{ALL} &{241}    &{121}    &102 (84.30\%)    &42.32\%\\
    \bottomrule
    \end{tabular}%
}
  \label{table_application}
\end{table}%

Table \ref{table_application} shows the results of enhancing HUAWEI's private security knowledge base, where column \textbf{\#HWTrees} indicates the original number of attack trees in their knowledge bases, \textbf{\#Accept} indicates the number of accepted trees, and \textbf{Increment Ratio} shows the increment of the knowledge base.
We can see that, 102 out of 121 synthesized attack trees are accepted with a ratio of 84.30\%, and the overall increment ratio for the entire database is 42.32\%. 
To sum up, {{\tool} helps with enhancing the private attack tree database.}

\begin{center}
\vspace{-0.2cm}
\begin{tcolorbox}[colback=gray!10,
                  colframe=black,
                  width=\columnwidth,
                  arc=1mm, auto outer arc,
                  boxrule=0.4pt,
                  left=0.1pt,
                  right=0.1pt,
                  top=0.1pt,
                  bottom=0.1pt
                 ]
\textit{\textbf{Answering RQ4}: {\tool} can be utilized to enhance the public and private security knowledge bases.
For the public knowledge bases CVE and CAPEC, 40.50\% and 53.72\% of them are earlier in Stack Overflow, and 16.53\% and 19.83\% are potentially new attacks.
For the HUAWEI's knowledge base, 84.30\% of the synthesized attack trees are accepted, and the entire database is incremented with 42.32\%.}
\end{tcolorbox}
\end{center}
\section{Human Evaluation for Attack Tree Synthesizing}\label{sec:human_evaluation}

In this section, we conduct the human evaluation of the quality of synthesized attack trees with its experienced practitioners.
First, we obtain the security posts that relate to the 121 new attack trees synthesized by {\tool} (Section \ref{sec:practical}). Then, we apply two representative baselines in the evaluation, i.e., Structured-Joint+ and GPT-3, on these security posts to synthesize the attack trees, where these DL and LLM models obtain the highest performances except for {\tool}.
After these two steps, we can separately obtain 121 attack trees for each baseline.
Third, we ask the practitioners of HUAWEI to manually inspect the attack trees synthesized by {\tool} and the two baselines, then rate these attack trees with the following two criteria:

\begin{itemize}[leftmargin=*]
    \item \textbf{{Accurateness}}: Whether the attack tree is accurate to describe a real attack in software security.
    \item \textbf{{Adequateness}}: Whether the attack tree is adequate to describe the details of attacks.
\end{itemize}
where we ask the practitioners to rate 1-5 under the above criteria. For each attack tree, a score of 5 means that the practitioner is satisfied with this attack tree, and a score of 1 means that the attack tree is not satisfactory.
Finally, the security practitioners send the feedback through internal email, and we gather each practitioner's feedback on these two criteria and present the distribution of scores in the violin plot.

\begin{figure}[t]
\centering
\vspace{-0.3cm}
\includegraphics[width=0.8\columnwidth]{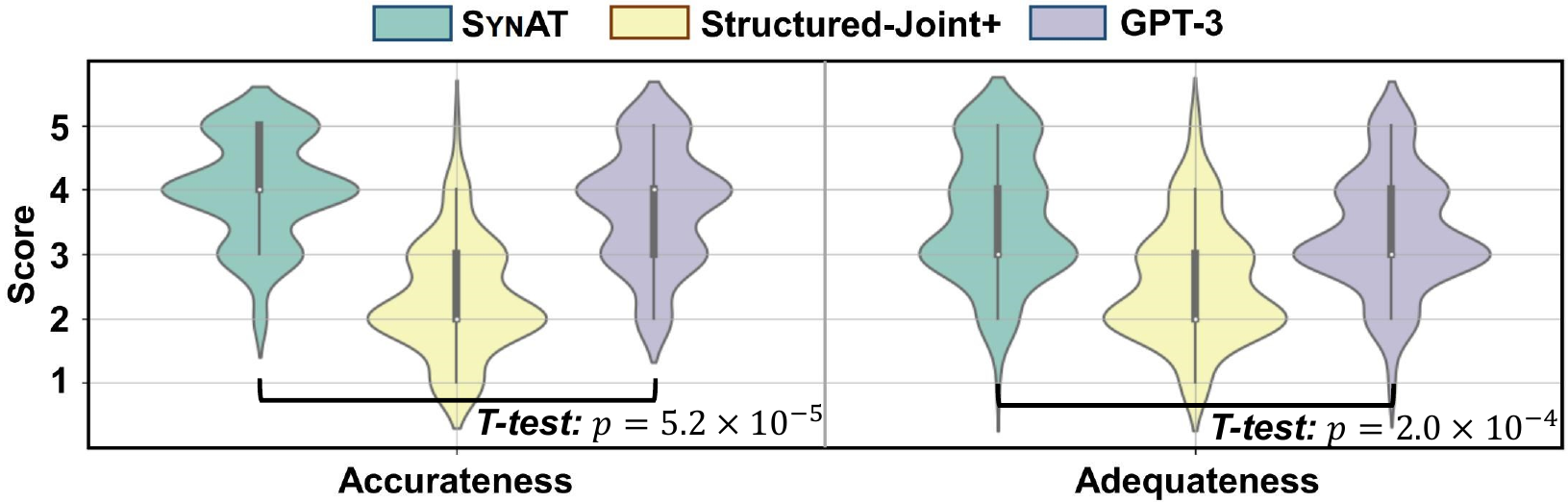}
\vspace{-0.2cm}
\caption{The results of human evaluation on attack tree synthesizing.}
\label{fig:human}
\end{figure}

{Fig. \ref{fig:human} shows the score distribution on the two criteria for {\tool}, Structured-Joint+, and GPT-3.
We can see that, 
the maximum distribution score of {\tool} is all over 3 on the three criteria, which is higher than the Structured-Joint+ and GPT-3. The average scores of {\tool} are 3.8, which is also higher than the Structured-Joint+ (2.4) and GPT-3 (3.5). These results indicate that practitioners are generally satisfied with the attack trees extracted by {\tool}.}
Moreover, we also utilize the significance testing to analyze the significance of {\tool}'s improvement to the GPT-3.
We choose the T-test with bilateral and independent samples to analyze the differences between {\tool} and GPT-3, which is widely used in significance testing. The result shows that the \textit{p-values} on accurateness and adequateness are $5.2\times 10^{-5}$ and $2.0\times 10^{-4}$, which means the scores of {\tool} are significantly higher than the SOTA LLM baseline, i.e., GPT-3\footnote{The value $p<0.01$ indicates the significant differences between two sets of scores.}.

Furthermore, we interview these security practitioners regarding the perceived usefulness of {\tool}, and they indicate that {\tool} is quite helpful. 
The practitioners demonstrate that some attacks, such as \textbf{format-string vulnerability} have not been included in the database, which is a common threat encountered in the coding process. 
{\tool} finds the attack trees with these missing attacks from the security posts, which can be used to replenish the security knowledge.  
Here is one of the comments:


\begin{center}
\vspace{-0.2cm}
\small
\begin{tcolorbox}[colback=white,
                  colframe=black,
                  width=\columnwidth,
                  arc=1mm, auto outer arc,
                  boxrule=0.4pt,
                  left=0.1pt,
                  right=0.1pt,
                  top=0.1pt,
                  bottom=0.1pt,
                colbacktitle=white!80!gray, coltitle=black, 
                 ]
\textit{One of the attack methods from the Format-String attack tree provided by {\tool} is `execute harmful code to attack buffer string in \texttt{printf(buffer)}'. 
When reviewing this, I just realized that my previous code is also vulnerable due to unsafely using the \texttt{printf} function. To make our colleges more cautious, I added this information to our security knowledge base.
} 
\end{tcolorbox}
\end{center}

\section{Discussion}\label{sec:discussion}

\subsection{Usage Scenarios for Designing Appropriate Security Patches}

{The attack tree can model the software risks that the program may encounter.
Compared with the textual description, the attack trees contain detailed attack methods and goals that indicate how the attackers harm the software system.
For developers unfamiliar with the current vulnerability they have encountered, the synthesized attack trees provide some cases with similar attack goals that can guide them in fixing their vulnerabilities.
Therefore, the cost of learning a new vulnerability has decreased with the help of synthesized attack trees from applying the {\tool}.

In addition to the previous practical applications, {\tool} has another practical usage scenario, i.e., designing the patches for vulnerable code in the Open Source Software (OSS).
Some OSS projects may contain vulnerable code, and if attackers utilize this unfixed vulnerable code, they may deploy exploits to harm the systems~\cite{DBLP:conf/sigsoft/PanZC0BHLH22}.
However, it is difficult for developers unfamiliar with software security to understand how the attackers may achieve the attacks and design appropriate patches to fix these vulnerabilities.
The synthesized attack trees provide historical information that describes how the attackers achieve the attack goals. This can guide the developers to analyze how the attackers may potentially exploit the vulnerabilities and design appropriate patches based on the attack methods~\cite{schneier1999attack}.
However, there is currently no large-scale public security knowledge base with cases of attack trees existing in the software community.
Although HUAWEI has built an attack tree database, it is private and has not been updated and maintained for a long time, as is illustrated in Section \ref{sec:practical}.
In comparison, the {\tool} is an automated approach that is fully trained on the security posts in the Stack Overflow and can be applied to synthesize attack trees from multiple online knowledge-sharing platforms.
Previous experiments show that the {\tool} can build a large-scale database that is utilized to enhance public/private security databases. 

\begin{table}[t]
\caption{{The results of applying {\tool} on designing the security patches for the vulnerable OSS projects.}}
\small
  \vspace{-0.2cm}
    \resizebox{\textwidth}{!}{
\begin{tabular}{c|ll|lll}
\toprule
\textbf{Types}             & \textbf{Approach} & \textbf{Training Strategy} & \multicolumn{1}{c}{\textbf{Fix@1}} & \multicolumn{1}{c}{\textbf{Fix@5}} & \multicolumn{1}{c}{\textbf{Fix@10}} \\ \midrule
\multirow{3}{*}{\textbf{Manual}}    & \textbf{Human}            & -                          & 20/37 (54.05\%)                    & 23/37 (62.16\%)                    & 24/37 (64.86\%)                     \\
                           & \textbf{+HWTrees}         & -                          & 22/37 (59.46\%)                    & 25/37 (67.57\%)                    & 28/37 (75.68\%)                     \\
                           & \textbf{+{\tool}\&HWTrees}  & -                          & \textbf{30/37 (81.08\%)}                    & \textbf{32/37 (86.49\%)}                    & \textbf{37/37 (100.00\%)}                    \\ \midrule
\multirow{4}{*}{\textbf{Automatic}} & \textbf{CodeT5}           & Fine-Tuning                & 15/37 (40.54\%)                    & 20/37 (54.05\%)                    & 20/37 (54.05\%)                     \\
                           & \textbf{+{\tool}}           & Fine-Tuning                & 25/37 (67.57\%)                    & \textbf{32/37 (86.49\%)}                    & 33/37 (89.19\%)                     \\
                            \cline{2-6}
                           & \textbf{ChatGPT}          & In-Context Learning        & 23/37 (62.16\%)                    & 24/37 (64.86\%)                    & 24/37 (64.86\%)                     \\
                           & \textbf{+{\tool}}           & In-Context Learning        & 29/37 (78.38\%)                    & 30/37 (81.08\%)                    & 31/37 (83.78\%)                     \\ \bottomrule
\end{tabular}}
\vspace{-0.3cm}
\label{tab:patch_design_application}
\end{table}

To evaluate the contribution of {\tool},
we first gather the vulnerability-related issue reports from the GHArchive~\cite{GHArchive}, which is a data source that achieves the latest vulnerability-related issue reports in the software community.
These vulnerability-related issue reports indicate the vulnerabilities in the corresponding OSS and the developers describe the details of how they observe these vulnerabilities in their projects, but some vulnerable code is not fixed after the issue report is released.
We have manually obtained 37 unfixed vulnerabilities from the GHArchive dataset after Jan 1st, 2024 based on the vulnerability-related issue reports, where some unfixed vulnerabilities may correspond to the attack trees synthesized by {\tool}.

Then, we experiment with two tasks, i.e., manual and automatic security patch designing.
\textbf{(1) For manual patch designing}, we ask the security practitioners in the HUAWEI to discuss with each other and manually design the patches for the 37 unfixed vulnerabilities, then they utilize the HUAWEI's attack trees (241 trees) as well as {\tool}'s synthesized attack trees (121 trees) in Table \ref{table_application} to help them design the patches;
\textbf{(2) For automatic patch designing}, since HUAWEI's attack tree database is private, we cannot directly use it in the automatic security patch generation, so the analysis of practical usage on automatic security patching is only conducted on the {\tool}'s synthesized 121 attack trees. 
We introduce two LLMs for code generation, i.e., CodeT5~\cite{DBLP:conf/emnlp/WangLGB0H23} and ChatGPT~\cite{LLMBackground}. These two LLMs are end-to-end text generation models that can analyze vulnerable code and recommend the appropriate security patches. Compared with other LLMs, fine-tuning the CodeT5 and utilizing in-context learning in ChatGPT can obtain the SOTA performances on fixing the vulnerable code.
In both manual and automatic patch designing, we obtain Top-$K$ patches that are most relevant to the vulnerability and calculate the fixing rate \textit{$Fix@K$}.
The metric is calculated with the following equation.
\begin{equation}
    Fix@K=\frac{\#(Trig\_Vul@K\cap Fix\_Vul@K)}{\#Total\_Vul@K}
\end{equation}
where “\#” is the symbol of the number calculation of evaluation samples. The number $\#Total\_Vul@K=37$, and $Fix@K=1$ if both the vulnerability triggering and fixing are satisfied in Top-$K$, i.e., $\#(Trig\_Vul@K\cap Fix\_Vul@K)=1$, and we make sure that the vulnerable code is triggered and the usefulness of the patch by using the vulnerability detection tools~\cite{zed, wapiti} and manual analysis.

Table \ref{tab:patch_design_application} shows the results of generating the patches for fixing the vulnerabilities with the help of attack trees, where \textbf{+HWTrees} indicates that using the HUAWEI's attack trees to guide the patch designing, and \textbf{+{\tool}} means using the attack trees generated by {\tool}. We can see that utilizing the attack trees can significantly improve the fixing rate $Fix@K$ of both manual and automatic patch designing.
For manual patch designing, introducing both HUAWEI\&{\tool}'s attack trees improves the fixing rate with +35.14\% (Fix@10), and the security practitioners accurately fix all 37 vulnerabilities after they refer to the attack trees.
For automatic patch designing, introducing the attack trees generated by {\tool} can also improve the Fix@10 with +25.14 (CodeT5) and +18.92 (ChatGPT).
In summary, introducing the {\tool} can effectively guide the ability to design appropriate patches for fixing the vulnerabilities.}

\subsection{Effectiveness of Restricting the Scope of Sentences}\label{sec:limiter}

To analyze the effectiveness of restricting the sentence scope, 
we compare the {\tool} with different types of prompt templates, and the number of selected sentences $K$, on event and relation extraction.

\subsubsection{Effect of Prompt Template}
{The design of the prompt template in LLM affects the performance of {\tool}. First, we compare the original \textbf{Cloze Template} with the Q\&A format \textbf{Prefix Template}~\cite{DBLP:conf/acl/LiL20}, which indicates the effect of template types. Then, we compare the template with the \textbf{Template without (w/o) Title}, which illustrates the effect of \textit{\{Title\}} in the template. The content of the two variants are shown as follows:

\begin{center}
\vspace{-0.2cm}
\small
\begin{tcolorbox}[colback=white,
                  colframe=black,
                  width=\columnwidth,
                  arc=1mm, auto outer arc,
                  boxrule=0.4pt,
                  left=0.1pt,
                  right=0.1pt,
                  top=0.1pt,
                  bottom=0.1pt,
                colbacktitle=white!80!gray, coltitle=black, 
                title={\textbf{Prompt Variants}}
                 ]
\textit{\textbf{Prefix-Template:} From the following security post \{Question, Accepted\_Answer\} (Title is \{Title\} to summarize the main topic of the security post), what are the restricted sentences that may contain the attack trees? \textbf{[Y]}} 
\textit{(Definition of attack trees.)}
\textit{\textbf{Task:} Please predict the token \textbf{[Y]} with $K$ restricted sentences from the security post's question and answers to answer this question. The format of the output is the bullet list with \textbf{Sentence 1} to \textbf{Sentence $K$}.}

\textit{\textbf{Template w/o Title:} From the following security post \{Question, Accepted\_Answer\}, \textbf{[Y]} are restricted sentences that may contain the attack trees.} 
\textit{(Definition of attack trees.)}
\textit{\textbf{Task:} Please predict the token \textbf{[Y]} with $K$ restricted sentences from the security post's question and answers to make the cloze-testing complete. The output format is the list with \textbf{Sentence 1...$K$}.}

\end{tcolorbox}
\end{center}



Fig. \ref{fig:classifier} (a) illustrates the comparison of these three templates. We can see that, the cloze template outperforms the prefix template on event extraction (+3.2\%) and relation extraction (+5.7\%), and it also outperforms the template w/o title on event extraction (+2.7\%) and relation extraction (4.5\%). 
In summary, the cloze template contributes to the {\tool} than other templates.}

\begin{figure}[htbp]
\centering
\vspace{-0.3cm}
\subfigure[Effect of prompt templates in restricting the scope of sentences.]{\includegraphics[width=0.35\columnwidth]{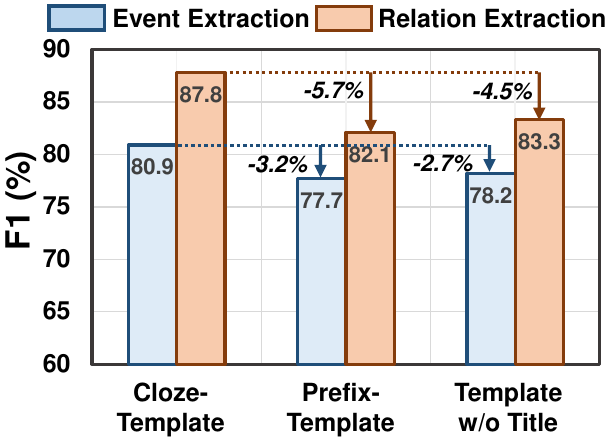}
\vspace{-0.3cm}
}
\hspace{1cm}
\subfigure[Effect of $K$ (number of selected sentences in restricting the scope).]{\includegraphics[width=0.35\columnwidth]{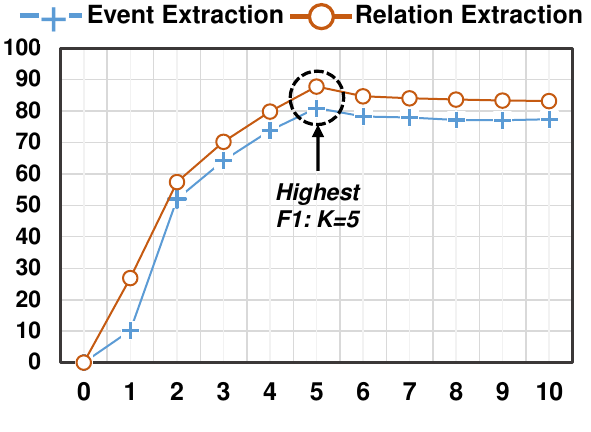}
\vspace{-0.3cm}
}
\vspace{-0.3cm}
\caption{The effect of restricting the scope of sentences.}
\label{fig:classifier}
\end{figure}

\subsubsection{Effect of Selected Sentences $K$.}
We analyze the effect of selected sentence $K$ on the two tasks F1 in $K\in\{1,2,...,10\}$.
Fig. \ref{fig:classifier} (b) illustrates the comparison results.
We can see that, when $K\in\{0,...,5\}$, the F1 performances of both event and relation extraction are largely increased, where event extraction increases from 0\% to 80.9\%, and relation extraction increases from 0\% to 87.8\%. When $K\in\{5,...,10\}$, the F1 performances of event and relation extraction decrease slightly by around 5\%.
In summary, the sentence-scope limiter contributes to the {\tool}, and {\tool} achieves the optimal value when $K=5$.

\subsection{Analysis of Unaccepted Attack Trees}
{
In the human evaluation, we find that 15.7\% (19/121) attack trees are not accepted by the analysts.
We manually inspect the unaccepted attack trees and find two reasons for their unacceptability: the \textbf{Duplicate} and \textbf{Low-readability} attack trees.}

\begin{itemize}[leftmargin=*]
    \item \textbf{Duplicate Attack Tree (10.7\%):} There are 10.7\% (13/121) synthesized attack trees that describe the attacks that have already existed in HUAWEI's security knowledge base. For example, the attack tree synthesized from post \#49168789 describes the server attacking from the XHR request, and we receive the feedback as \textit{{``This attack tree accurately describes how the XHR request happens. However, it has already been included in our database: T-Tampering/ServerAttack, so we did not adopt it''.}} From the feedback, we can see that, although the synthesized attack tree is fine, it has already been included in the company's database. To address this issue, we plan to add a module to analyze the similarity between the extracted attack tree and the attack trees in the database, thus excluding the duplications.
    \item \textbf{Low-readability Attack Tree (5.0\%):} There are 5.0\% (6/121) synthesized attack trees that are considered hard to understand by security analysts. Since the developer's discussions on Stack Overflow are typically informal, they are inevitable to contain some \textit{{jargon}} or \textit{vague expressions}. If contained in our extracted results, it will lower the readability of the attack tree. To address this issue, we plan to add a coreference resolution module~\cite{DBLP:journals/ai/LuLTHS22} to analyze the meaning of jargon. We also plan to add a perplexing measurement to reduce the vague attack methods.
\end{itemize}



\subsection{Threats to Validity}

In this section, we aim to discuss the internal, external, and constructive threats of {\tool} and how to mitigate these threats in our approach.

\textbf{Internal Threats.}
{The internal threat comes from the experiments. Due to the lack of labeled data, the performance rule-based attack tree construction is difficult to evaluate. To alleviate this, 
we evaluate the practical application of {\tool}, where 84\% of attack trees synthesized by rules are accepted. 
The result indicates that rule-based construction can help build better attack trees.}
There is also a threat coming from our application study. 
{We cannot guarantee that the evaluations of security analysts on attack trees are all fair. 
To mitigate this threat, 
{we conduct the training course, thoroughly discussing {\tool} and the extracted attack trees with all the security practitioners, which reduces the occurrence of subjective evaluation to some extent.}}

\textbf{External Threats.}
The external threat comes from the quality of the dataset from the security posts, as well as the data augmentation strategy we use. 
In this study, we augment the dataset from 5,070 posts to 18,203 posts, which may lead to the augmented samples having semantic differences from the original ones. {We manage this threat by manually inspecting 200 augmented samples. It turns out that 91\% of the inspected samples are correct, and only 9\% of samples have slight differences.
}

\textbf{Constructive Threats.}
The constructive threat lies in the metrics used in our study. We use the commonly used metrics (Precision, Recall, and F1) to evaluate the performance, and the metrics (AHD and TEDS) to evaluate the performance of the synthesized attack trees.
We also manually label the attack events and relations as the ground truth when calculating the performance metrics. This threat can be largely relieved as all the instances are reviewed with a concluding discussion session to resolve any disagreement in labeling. 




\section{Related Works}
\label{sec:related}

In this section, we discuss the related works of {\tool}. The prior studies relate to manually constructing attack trees and analyzing security online discussions.

\textbf{Manually Constructing Attack Trees.}
{Many works have recently been proposed to construct attack trees to analyze system risks and defend against attacks. Most of them utilize template or rule-based syntax analysis to construct attack trees based on structural data sources.}
Traditionally, the attack trees are manually built based on the dependencies, such as Goal-inducing Attack Chains (GACs) and attack graphs~\cite{DBLP:conf/iwia/DawkinsH04,DBLP:conf/csfw/VigoNN14,DBLP:journals/corr/Paul14,DBLP:conf/csfw/IvanovaPHK15,DBLP:books/sp/18/KarrayDGE18,DBLP:conf/stm/JhawarLMR18}.
Recently, the attack trees have been constructed based on the abstract description of workflow and identification of hazards. Xu et al. \cite{DBLP:conf/syscon/XuVS16} constructed attack trees by transforming fault trees.
Lemaire et al. \cite{DBLP:conf/critis/LemaireVDN17} utilized templates to construct attack trees for evaluating the security of cyber-physical systems.
Mentel et al. \cite{DBLP:conf/csfw/MantelP19} constructed an attack tree from a set of generated implications for the given attack goal by backward chaining.
Pinchinat et al. \cite{DBLP:conf/gramsec/PinchinatSC20} proposed a model-free method by taking an abstract model for some expert knowledge as input. 
Kern et al. \cite{DBLP:conf/isse2/KernLB021} proposed a preliminary system architecture model consisting of hardware network architecture, logical functional architecture with data dependencies, and threat agents to construct attack trees. 
{Different from the previous work, we focus on automatically constructing attack trees from security posts of Stack Overflow based on event and relation extraction, which can extract valuable security information from knowledge-sharing platforms and facilitate secure software development.}

\textbf{Analyzing Security Online Discussion.} 
Recently, more and more developers leveraged open-source community platforms, e.g., GitHub, Stack Overflow, and Security StackExchange, to post their security-related concerns and discuss solutions with other developers. These large-volume security data contain rich information for understanding security discussions.
Some researchers, such as Pletea et al. \cite{pletea2014security} investigated the atmosphere and mood of security discussions on Github, 
finding that developers showed more negativity in security discussions than in non-security discussions. 
Others studied the topic of security discussions and analyzed the area that most security practitioners care about~\cite{meyers2019pragmatic,le2021large,yang2016security,zahedi2018empirical}.
Among them, Le et al. \cite{le2021large} and Yang et al. \cite{yang2016security} utilize the automatic approach, such as Latent Dirichlet Allocation (LDA) to analyze the discussion topics on Stack Overflow and Security StackExchange.
Zahedi et al. \cite{zahedi2018empirical} used an approach that combined topic modeling with qualitative analysis to summarize 26 security-related topics on GitHub, 
Our work is different from the previous works as we focus on extracting attack trees from these security conversations, and aim to increase the opportunity to enrich the security database and develop secure software by reusing this attack knowledge.

\section{Conclusion and Future Work}
In this paper, we proposed {\tool}, an automated approach to synthesize attack trees from online discussions on Stack Overflow. 
{\tool} first restricts the sentence scope with the LLM, which may contain the attacks from the security post; then {\tool} utilizes a transition-based event and relation extraction model to extract the events and relations simultaneously;
finally, {\tool} constructs the attack trees with a set of heuristic rules from the extracted events and relations.
{We conducted experiments on 5,070 security posts collected
from Stack Overflow and compared {\tool} with multiple representative baselines on event extraction and relation extraction.}
The experimental results show that our approach outperforms all the baselines on event and relation extraction. 
{\tool} was also successfully applied to enhance HUAWEI's security knowledge base and public security knowledge base.

In the future, we plan to further enhance the {\tool} with more extended datasets from other crowd security platforms, such as GitHub Issue Reports (IRs). We also plan to combine the rich-text information, such as code blocks and page screenshots, in the extraction and refine the synthesized attack tree with mitigations.
\bibliographystyle{ACM-Reference-Format}
\bibliography{ref}

\end{document}